\documentclass[aps,10.5pt,notitlepage,nofootinbib,superscriptaddress]{revtex4}

\usepackage{todonotes}
\usepackage{hyperref}
\usepackage{amsfonts,amssymb,amsmath,exscale,bbm}
\usepackage[all,knot]{xy}  
\usepackage{color}
\usepackage{graphicx}
\usepackage{subfig}

\usepackage{amsmath}
\usepackage{amsfonts}
\usepackage{graphicx}
\usepackage{psfrag}
\usepackage{array}
\usepackage{bbm}
\usepackage{hyperref}
\usepackage{amsthm}

\theoremstyle{definition}

\newcommand\beq{\begin{equation}}
\newcommand\eeq{\end{equation}}
\newcommand{\be}{\begin{equation}}
\newcommand{\ee}{\end{equation}}
\newcommand{\bes}{\begin{eqnarray}}
\newcommand{\ees}{\end{eqnarray}}

\newcommand\beqa{\begin{eqnarray}}
\newcommand\eeqa{\end{eqnarray}}

\def\tl{\widetilde}

\def\la{\langle}
\def\ra{\rangle}

\newcommand{\SU}{\mathrm{SU}}

\newcommand\acts\triangleright

\begin{document}


\title{\bf Group field theory as the 2nd quantization of Loop Quantum Gravity}

\author{{Daniele Oriti}}
\affiliation{Max-Planck-Institut f\"ur Gravitationsphysik (Albert Einstein Institute) \\ Am M\"uhlenberg 1, D-14476 Golm, Germany, EU \\ email: daniele.oriti@aei.mpg.de}

\date{\small\today}

\begin{abstract}
We construct a 2nd quantized reformulation of canonical Loop Quantum Gravity at both kinematical and dynamical level, in terms of a Fock space of spin networks, and show in full generality that it leads directly to the Group Field Theory formalism. In particular, we show the correspondence between canonical LQG dynamics and GFT dynamics leading to a specific GFT model from any definition of quantum canonical dynamics of spin networks. We exemplify the correspondence of dynamics in the specific example of 3d quantum gravity. The correspondence between canonical LQG and covariant spin foam models is obtained via the GFT definition of the latter. 
\end{abstract}

\maketitle


\section{Introduction}
\label{sec:intro}
Group field theories \cite{GFT,GFTthomas} are a generalization of matrix models and an enrichment of tensor models \cite{tensorreview} through the addition of group-theoretic data interpreted as pre-geometric data, \lq seeds\rq ~of the continuum geometry that these models should generate in a continuum approximation. Indeed, they have been first introduced \cite{boulatov} in parallel to tensor models \cite{tensor} in the early 90's in the context of topological field theories and in particular to give a tensorial formulation to lattice models like the Ponzano-Regge model and its 4d generalization. The Ponzano-Regge model and its higher dimensional extensions were linked, soon afterwards \cite{carloPR}, to canonical loop quantum gravity \cite{LQG}, in that their boundary states were of the same type: spin networks. Indeed, they are now seen as the prototype of spin foam models \cite{SF}, introduced in the same period as a covariant formalism with the potential to provide a complete definition of loop quantum gravity dynamics, as an algebraic and combinatorial version of the sum-over-geometries idea. The impetus for the development of group field theories has come in fact from the spin foam corner of loop quantum gravity \cite{DP-F-K-R, P-R, laurentGFT, tenGFT}, over the last decade\footnote{Even more recently, the resurgence of tensor models as a powerful definition of random discrete geometry has provided further stimulus to the group field theory approach, by providing several key analytic tools to control the sum over cellular complexes that is the combinatorial backbone of group field theories.}. The basic relation between group field theories and spin foam models has been clarified early on \cite{mikecarlo}. We now know that there exist a one-to-one correspondence between spin foam models and group field theories, in the sense that for any assignment of a spin foam amplitude for a given cellular complex, there exist a group field theory, specified by a choice of field and action, that reproduces the same amplitude for the GFT Feynman diagram dual to the given cellular complex\footnote{We are glossing over the issue of the choice of combinatorial structures for boundary states as well as bulk complexes. Usually, these are chosen to be simplicial complexes, which entails specific restrictions on the valence of spin network graphs and spin foam vertices. This is matched by restrictions on the GFT field content and combinatorics of field arguments in the GFT interactions. Loop Quantum Gravity does not suggest this type of restrictions. This motivated well-defined but not entirely (geometrically) justified generalizations of known spin foam models to general combinatorial structures \cite{KKL}. The reason we gloss over this point is that it is rather straightforward to see \cite{JohannesJimmyDaniele} that group field theories can be constructed also for these combinatorial generalizations, even if at the cost of the same type of loss of computational power.}. Conversely, any given group field theory is also a definition of a spin foam model in that it specifies uniquely the Feynman amplitudes associated to the cellular complexes appearing in its perturbative expansion. Thus group field theories encode the same information and thus define the same dynamics of quantum geometry as spin foam models. This is the basic fact. However, a stronger claim can be justified. Unless one believes that a fundamental theory of quantum spacetime possesses a finite number of degrees of freedom, it is clear that a spin foam formulation of quantum gravity cannot be based on a single cellular complex. A complete definition should involve an infinite class (to be better identified) of such cellular complexes, in the same way in which the Hilbert space of the theory is defined over an infinite class of spin network graphs. The very definition of a spin foam model, therefore, entails a prescription for organizing the amplitudes associated to all such complexes. This organizing prescription can in general take the form of some \lq\lq refinement\rq\rq of spin foam complexes \cite{SFrefinement} or of \lq\lq summing\rq\rq over them. Group field theory provides one such prescription as it generates a sum over complexes, weighted by spin foam amplitudes, as a Feynman diagram expansion and thus with canonically assigned combinatorial weights. Thus one can say that group field theory is not only equivalent to spin foam models, but it is actually a {\it completion} of the spin foam definition of the quantum dynamics of spacetime. 

The exact correspondence between spin foam models and canonical loop quantum gravity is not yet fully understood, despite many results \cite{LQG-SF}, beyond the general idea of them being a covariant versus canonical formulation of a theory of spin networks. The equivalence between spin foam models and group field theories proves that a strict relationship should exist between the latter and canonical loop quantum gravity. In fact, the idea that group field theories are a second quantized version of loop quantum gravity has been voiced repeatedly in the past \cite{GFT,tenGFT,laurentGFT}, even though the details of such correspondence have not been clarified or detailed, leaving room for doubting that such correspondence actually exists. 

In this paper we show that not only this correspondence exists, but that it is indeed the result of a straightforward second quantization of spin networks kinematics and dynamics, which allows to map any definition of a canonical dynamics of spin networks, thus of loop quantum gravity, to a specific group field theory encoding the same content in field-theoretic language. This map is very general and exact, on top of being rather simple. It puts in direct correspondence the quantum states of the canonical theory and its algebra of quantum observables, including any operator defining the quantum dynamics, with a GFT Fock space of states and algebra of operators (constructed out of fundamental field operators), and its dynamics, defined in terms of a classical action and quantum equations for its n-point functions.

In order to show this correspondence, one does not need to pass through the spin foam definition of the same dynamics, which, for reasons that will become clear, would actually make establishing the correspondence less straightforward. On the contrary, the correspondence so established {\it implies} and defines the correspondence between canonical LQG formulation and spin foam dynamics, which is then better understood via the GFT/LQG correspondence.

We will also argue that the 2nd quantization leading to GFT straight from canonical LQG \footnote{A 2nd quantized structure for GFT has been also sketched in its general features in \cite{mikovic}, without relating it to canonical LQG.} is particularly useful, in that it provides several key tools and a fresh perspective, and opens new directions to develop the theory further.

\subsection{The Hilbert space of Loop Quantum Gravity}
Before describing the second quantization of spin networks states and the correspondence between loop quantum gravity and group field theory, let us recall briefly what is the LQG Hilbert space and how one arrives at it (see \cite{LQG} for more details). This will help clarifying in what sense the GFT Hilbert space contains the same type of states in a second quantised form, but also how it differs from the standard construction.

The kinematical Hilbert space of loop quantum gravity (without matter degrees of freedom) carries a representation of the holonomy-flux algebra obtained by quantizing the classical Poisson algebra of holonomies of the Ashtekar connection and fluxes of the conjugate triad field. Coming from the canonical quantisation of GR in the continuum, this involves the holonomies associated to {\it all} paths embedded in the canonical hypersurface and the fluxes across {\it all} surfaces similarly embedded. Still, this vast set of degrees of freedom can be equivalently characterised by considering all {\it graphs} $\gamma$ embedded in the spatial manifold, the holonomies along the edges of these graphs, and the set of surfaces dual to them (i.e. such that each surfaces is pierced by one and only one edge of the graph).

To each such graph, one can associated a Hilbert space of states of quantum geometry. 
Each graph-based Hilbert space $\mathcal{H}_\gamma$ can be given in the connection representation as an $L^2$ space of cylindrical functions over a group manifold $G$\footnote{One could generalize this definition, and, actually, all the construction discussed in this paper, to homogeneous spaces $G/H$, without much trouble; this generalization could be useful in the context of 4d quantum gravity models \cite{SF, GFT-holst,GFT-BC}.} (usually, $SU(2)$) or, equivalently, in the flux representation \cite{flux,GFT-nc} as non-commutative functions on the Lie algebra of the same group. Moreover, each graph-based state is assumed to be gauge invariant; that is, one imposes a gauge invariance under simultaneous translation (left or right, depending on the orientation of the edges) of the group elements associated to edges of the graph, by a single group element associated to the vertex to which the same edges are incident; in the flux representation, the same condition is imposed by including in the definition of the states a non-commutative delta function for each vertex, imposing that the fluxes associated to the incident edges sum to zero\footnote{The spin network vertex can be interpreted \cite{spinnetPolyhedra} as dual to a polyhedron with the (non-commutative) fluxes associated to its boundary faces \cite{flux}. The gauge invariance condition represents then the closure of this polyhedron. The polyhedra are imagined as \lq glued\rq to one another across shared faces, thus with their boundary fluxes identified up to the opposite signs (due to the opposite orientation), when their dual vertices are connected by a link. Notice that the graph does not determine uniquely the topology of the corresponding cellular complex, in particular the shape of the faces dual to the links. This deficiency could be cured by suitable coloring of the same graphs, using techniques introduced in the context of group field theories and tensor models \cite{coloring}.}. This means that there exist a realization of the graph-based Hilbert space as $\mathcal{H}_\gamma = L^2\left( G^E/G^V\right)$. We work with directed graphs, with the orientation of each edge $(ij)$ between vertices $i$ and $j$ reflected in the assignment of group (Lie algebra) elements to it as: $g_{ij}= g_{ji}^{-1}$ ($X_{ij} = - X_{ji}$).

\

Considering the Hilbert spaces associated to all possible graphs embedded in the spatial manifold, one has 

$$ \bigcup_{\gamma} \mathcal{H}_\gamma \qquad . $$ This is of course itself not a Hilbert space. In order to turn it into a Hilbert space one could simply take the direct sum over the set of graphs defining the union, to get: 

$$ \mathcal{H}_1\,=\,\bigoplus_\gamma \, \mathcal{H}_\gamma \qquad .$$ In fact, just like the individual Hilbert spaces $\mathcal{H}_\gamma$ would be well-defined also in an abstract, non-embedded context, the full space $\mathcal{H}_1$ would be maybe the natural choice in absence of any embedding and of any continuum interpretation of the states. On the other hand, one can take advantage of the continuum embedding of the graphs and of the continuum interpretation of the algebraic data labelling them, to define a different Hilbert space including all possible graphs. One imposes appropriate {\it cylindrical equivalence} relations $(\approx)$ between quantum states associated to different graphs, reflecting properties of the underlying continuum connection field. These properties allow also to define a scalar product for states associated to different graphs, by embedding both into a common, larger graph $\gamma$ and using the scalar product of the corresponding Hilbert space $\mathcal{H}_\gamma$. The construction \cite{LQG, flux} results in the Hilbert space 
$$
\mathcal{H}_2 \, =\, \frac{\bigcup_\gamma \, \mathcal{H}_\gamma}{\approx} \qquad .
$$
One can show that this second Hilbert space can be given as a direct sum of graph-based Hilbert spaces, like $\mathcal{H}_1$, but of course with the individual graph-based Hilbert spaces different from $\mathcal{H}_\gamma$ (in essence, they correspond to the former Hilbert spaces \lq\lq without the zero modes\rq\rq). More importantly, one can also show that the same Hilbert space is also the result of a proper {\it continuum limit} of the theory defined by each $\mathcal{H}_\gamma$, realised via a projective limit of graph structures: $\mathcal{H}_2 \, =\, \lim_{\gamma \to \infty}\,\mathcal{H}_\gamma$, intuitively understood as the limit of \lq\lq infinitely refined graphs\rq\rq.

\

In the following we will consider a different Hilbert space for states of quantum geometry, and show that it contains all graph-based states forming  $\cup_\gamma \mathcal{H}_\gamma$. Then we will show that, on the one hand, it embeds faithfully, i.e. maintaining the vector space structure, each $\mathcal{H}_\gamma$, while, on the other hand, it organises differently states associated to different graphs, making them not necessarily orthogonal. Finally, we will show how this new Hilbert space admits a straightforward 2nd quantisation procedure turning it into a Fock space, easily seen to be the one underlying the GFT framework. We will restrict ourselves to graphs of {\it fixed valence} $d$. Both the 2nd quantization construction to be presented and the whole GFT formalism can be straightforwardly generalized to the full LQG Hilbert space \cite{JohannesJimmyDaniele}, at the cost of some formal complication that is not essential for the points we want to make in the present work.

\

This new Hilbert space may be seen as another natural way of organising graph-based states into a Hilbert space structure, alternative to $\mathcal{H}_1$ mentioned above, which does not make any use of continuum notions, and makes perfect sense in an abstract, non-embedded context. In fact, we do not impose any additional {\it cylindrical consistency} (or equivalence) condition on our quantum states, that would be instead necessary for a continuum interpretation in terms of a (generalized) connection field. A general, conceptual motivation is that we believe the continuum limit of the theory is to be realized {\it dynamically}, and thus we are reluctant to impose additional kinematical conditions that are not required by the dynamics, and may actually even interfere with such dynamical realization of continuum (quantum) geometry. This is just a philosophical bias, but it seems supported also by recent results obtained in the context of the coarse graining of spin foam models \cite{biancaCoarse}, where it is shown that a different type of cylindrical consistency conditions on boundary states is induced by such coarse graining of bulk quantum amplitudes \cite{biancaCylindrical}. Also, remaining at the pure kinematical level, various technical difficulties (fundamentally due to the non-abelian nature of the group $G$) have been identified in the attempt to use the counterpart of the usual cylindrical consistency conditions on connections in the flux representation (or, for what matters, in the spin representation), to define a representation of the same Hilbert space as an $L^2$ space of continuum generalized triad field \cite{FluxProjective}. Also these results seem to suggest that some generalization (more \lq\lq phase space based\rq\rq) of the usual cylindrical consistency conditions is called for. Finally, cylindrical equivalence is not compatible with the {\it restriction to fixed valence} $d$ for the graphs. 
We do however impose an additional restriction on the quantum states of the theory we consider; we require these to be invariant under graph automorphisms, which in particular includes vertex relabeling. This is a rather natural condition implying that the states depend only on the intrinsic combinatorial structure of the graphs themselves and can also be seen as a discrete counterpart of continuum diffeomorphisms. We will discuss it a bit more in the following, when the consequences of this symmetry requirement will be presented. 

\section{Spin network states as many-body quantum states}
The simple but key point of the construction is to realize in which sense LQG states (which we call here generically \lq spin network states\rq, even if this name would only strictly apply to LQG states in the spin representation) can be understood as \lq\lq many-particle\rq\rq states analogously to those found in particle physics and condensed matter theory.  

\begin{figure}[h]
\includegraphics[scale=0.3]{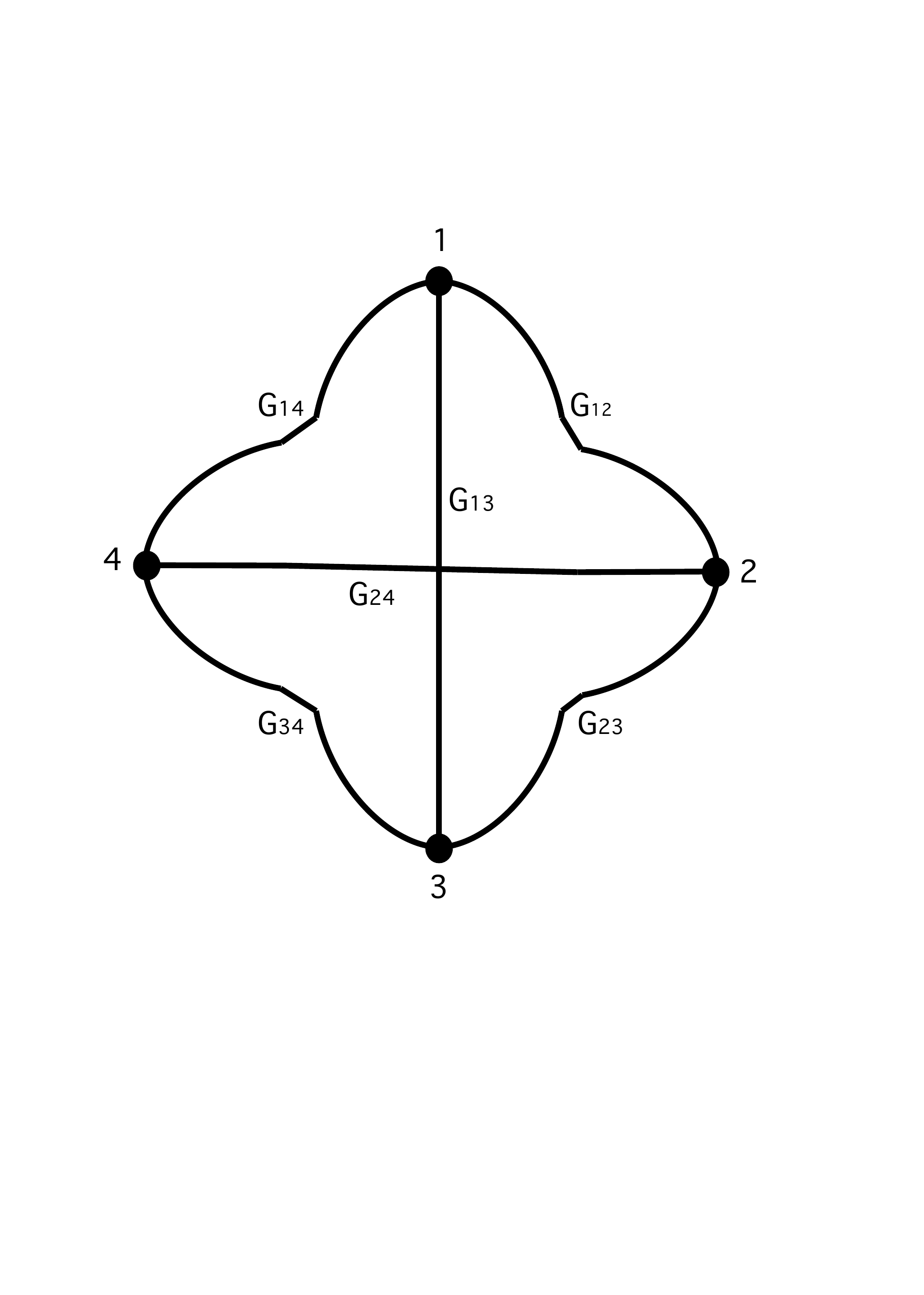} \hspace{-1cm}
\includegraphics[scale=0.3]{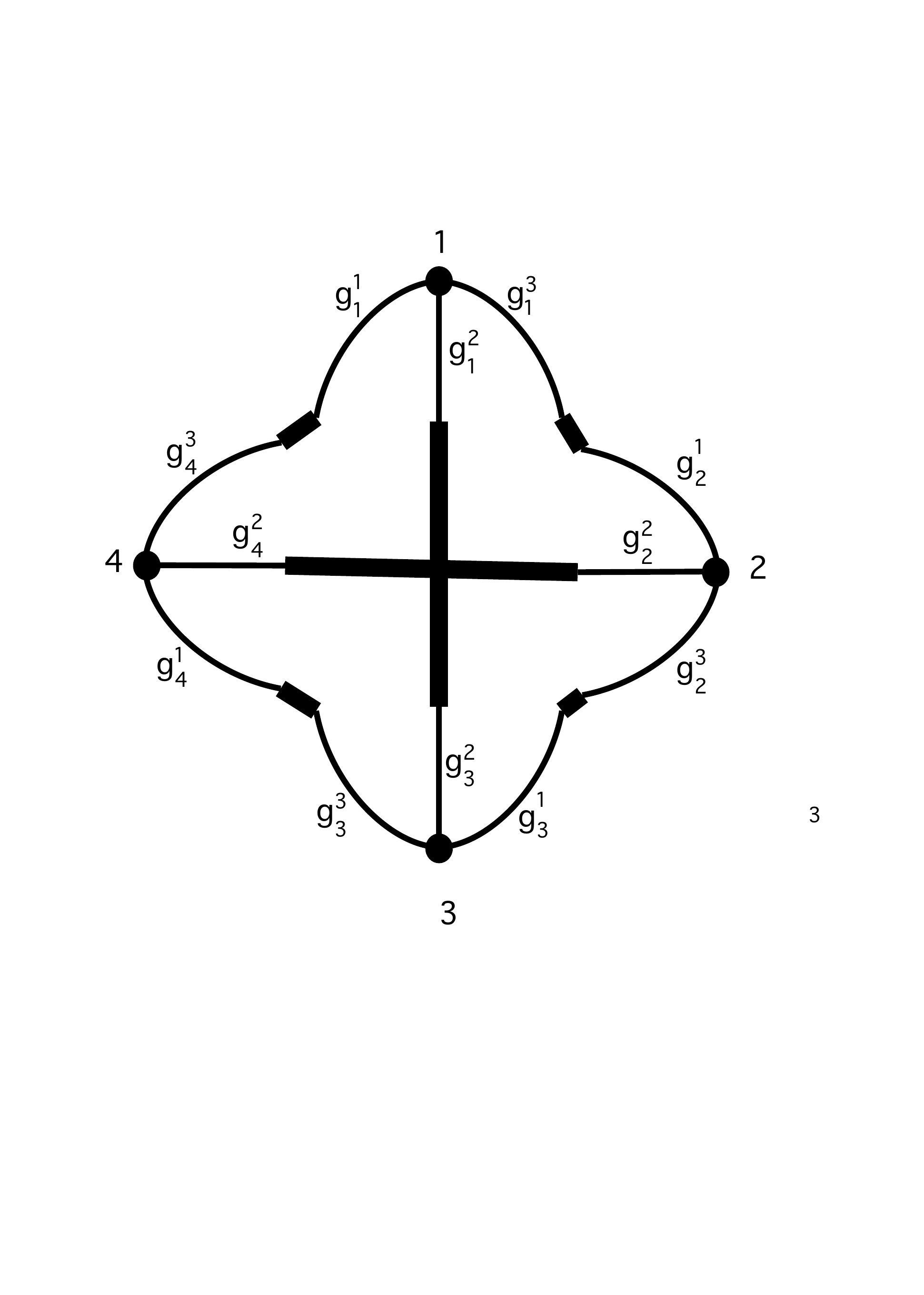} \hspace{-1cm}
\includegraphics[scale=0.3]{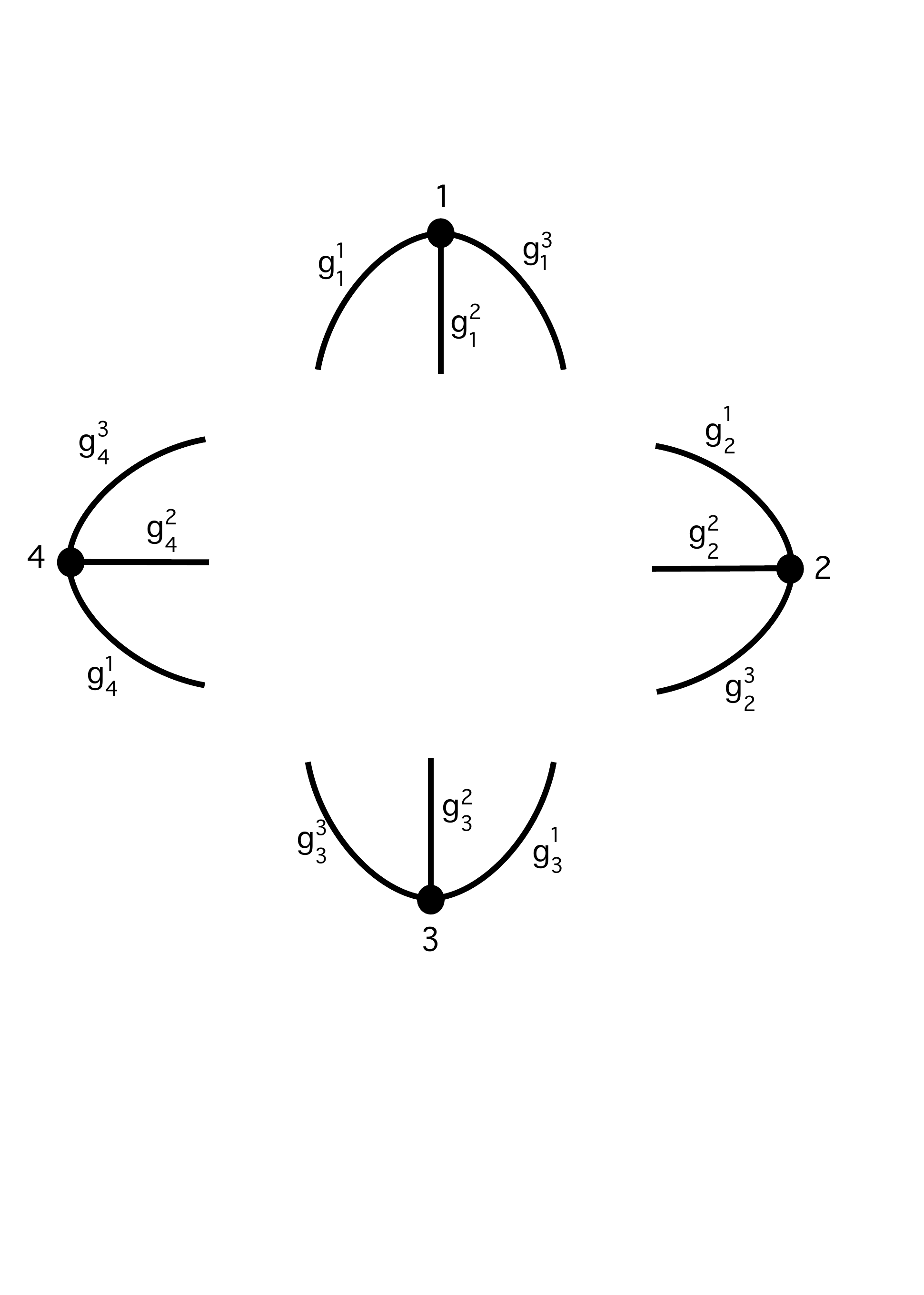}
\caption{A simple spin network (left) obtained from the gluing (center) of four spin network vertices (right). No need for the indices $(a,b,..)$ in this special case.}
\label{SNgluing}
\end{figure}

Consider a closed graphs with $V$ vertices, each assumed to be d-valent, i.e. with $d$ outgoing links from it. We label the vertices with the index $i=1,..,V$. We indicate the set of edges of the graph as $E(\gamma)\subset (\{1,\ldots,V\}\times\{1,\ldots,d\})^2$ (satisfying $[(i\,a)\,(i\,a)]\not\in E(\gamma)$). The connectivity of such a graph is specified as follows: if $[(i\,a)\,(j\,b)]\in E(\gamma)$, then there is a directed edge connecting the $a$-th link at the $i$-th node to the $b$-th link at the $j$-th node, with source $i$ and target $j$.
Consider now a generic cylindrical function
$\Psi_\gamma(G_{12}^{11},G_{13}^{21},..)=\Psi_\gamma(G_{ij}^{ab})\in \mathcal{H}_\gamma$ associated to the same graph, where the group elements $G_{ij}^{ab}\in G$ ($G = \SU(2)$ in LQG) are assigned to each link $e:=[(i\,a)\,(j\,b)]\in E(\gamma)$ of the
graph. 
As an example, one can consider the tetrahedral graph formed
by four vertices and six links joining them pairwise (see figure ~\ref{SNgluing}).
One has $G_{ij}=G_{ji}^{-1}$, and we
assume the gauge invariance: $\Psi_{\gamma}(G_{ij}) = \Psi_\gamma(\beta_i
G_{ij}\beta_j^{-1}) \quad \forall \beta_i \in G$ at each vertex $i$ of the graph. This is the type of functions forming the LQG Hilbert space.

Consider now a different Hilbert space $\mathcal{H}_V\,\simeq L^2\left( G^{d \times V}/ G^V \right)$. A generic element of this space will be a function of $d \times V$ group elements:
\be
\varphi(g_i^a)=\varphi(g_1^1,g_1^2,...,g_1^{d},.....,g_V^1,g_V^2,...,g_V^d) \;\;\; \label{varphi}.
\ee We again assume gauge invariance at the vertices in $V$ of the
graph: $\varphi(...g_i^a,...,g_j^b)=\varphi(...\beta_i
g_i^a,...,\beta_j g_j^b) \quad \forall\beta_i\in G$. As in the LQG case, we take the measure of such Hilbert space to be the Haar measure.
We associate each such function to a d-valent graph formed by $V$ disconnected components, each corresponding to a single d-valent vertex and $d$ 1-valent vertices, thus having $d$ edges (see figure ~\ref{SNgluing}). We refer to this type of disconnected components as {\it open spin network vertices}. 

Given a closed $d$-valent graph with $V$ vertices (specified by $E(\gamma)$), a cylindrical function $\Psi_\gamma$ can be obtained by group-averaging of a wavefunction $\varphi$,
\be
\Psi_\gamma(\{G^{ab}_{ij}\})\,=\,
\prod_{[(ij)(ab)]\in E(\gamma)}\int_{G}\,d\alpha_{ij}^{ab}\,
\phi(\{g_{i}^{a} \alpha_{ij}^{ab}; g_{j}^{b}\alpha_{ij}^{ab} \})
\,=\,\Psi_\gamma(\{ g^{a}_{i}(g_{j}^{b})^{-1} \})\,, \label{gluingGroup} \ee
in such a way
that each edge in $\gamma$ is associated with two group
elements $g_i^a,g_j^b\in G$. 

The integrals over the $\alpha$'s in the formula above operate a
\lq\lq gluing\rq\rq of the open spin network vertices corresponding to the function $\varphi$, pairwise along
common links (which ones end up being \lq glued\rq ~depends on which arguments of $\varphi$ we decide to impose the projection enforced by the integral over $\alpha$), thus forming the closed spin network represented by
the closed graph $\gamma$. The gluing can be
interpreted as a simple symmetry requirement, imposing that the
function $\varphi$ depends on the group elements $g_i^a$ and
$g_j^b$ only through the combination
$g_i^a(g_j^b)^{-1}\approx G_{ij}^{ab}$ (invariant under group action, by the {\it same} group element, at the endpoint of the two open edges to which these group elements are associated).

This can be taken as a -definition- of the function $\Psi$, as obtained from functions $\varphi \in\mathcal{H}_V$. In other words, the formula and the construction show that, only using functions $\varphi$, one can always construct a generic function $\Psi$ with all the right variables and symmetry properties.
In other words, we have shown that the space of the functions $\Psi$ is a subset of the space of functions $\varphi$. See again figure ~\ref{SNgluing} for an example of this construction.
Obviously, one can define states corresponding to composite, but still open graphs, formed by gluing several open spin network vertices together, but still presenting some open edge.

We will later use the notation $\{g_i^a\}=\vec{g}_i$ for labelling the
$d$ group elements attached to the d-valent vertex $i$, as arguments of the function $\varphi$ in $\mathcal{H}_V$ (with the proviso that the \lq arrow\rq does not indicate an underlying vector space). In using this notation, we assume an ordering of the group elements associated to a given open spin network vertex, which is chosen arbitrarily at this stage and kept fixed.

\medskip

There is nothing special, of course, about the connection (group) representation of the Hilbert space, and the same construction can be phrased entirely along the same lines in the flux representation and in the spin representation (or any other variation of the latter, like the spinor and the coherent state representations \cite{spinor}).

In the flux representation the states $| \varphi\rangle$ are non-commutative functions on $d \times V$ copies of $\mathbb{R}^D$ (where $D$ is the dimension of the group), i.e. the Lie algebra of $G$ seen as a vector space, that is one Lie algebra element for each edge of each connected component of the graph (each open spin network vertex), and subject to the closure condition at each vertex in $V$:

\be
\varphi(X_i^a)=\varphi(X_1^1,X_1^2,...,X_1^{d},.....,X_V^1,X_V^2,...,X_V^d) \equiv \varphi(\vec{X}_1,...,\vec{X}_V) = \int [dg_i^a] \varphi(g_i^a) \;\; \prod_{i=1}^V\delta_\star(X_i^1 + ... + X_i^d) \star \prod_{a=1}^d E_{g_i^a}(X_i^a)
\ee

and multiply via a $\star$-product whose explicit expression depends on the quantization map chosen for quantizing the classical flux variables \cite{FluxQuantizationMaps}. From the choice of quantization map depends also the expression for the plane waves $E_g(X)$ defining the non-commutative Fourier transform, which intertwines the group and the flux representations. 

Similarly the functions associated to closed graphs can be written in the flux representation as:

\be
\Psi_\gamma(X_{ij}^{ab}) = \int [dG_{ij}^{ab}] \Psi_\gamma(G_{ij}^{ab}) \prod_{[(ij)(ab)]\in E} E_{G_{ij}^{ab}}(X_{ij}^{ab})
\ee

Again, cylindrical functions associated to closed graphs can be obtained as a particular case of functions in $\mathcal{H}_V$, with the gluing condition for open spin network vertices now imposing the identification (up to orientation) of flux variables on the two edges 
being glued, the flux variables being associated to the boundary faces of the polyhedron dual to the spin network vertex:

\be
\Psi_\gamma(X_{ij}^{ab}=X_i^a=-X_j^b)\,=\,\prod_{[(ij)(ab)]}\,\varphi(...,X_i^a\,,...,X_j^b,...)\,\star\,\delta_\star(X_i^a + X_j^b)
\ee 
This formula is simply the same formula \ref{gluingGroup} in flux variables, as resulting from direct non-commutative Fourier transform \cite{flux}. Notice that the resulting state is still well-defined as an element of $\mathcal{H}_\gamma$ (as well as of $\mathcal{H}_V$) due to the properties of the non-commutative delta function, which is a square integrable function when seen as a function on $\mathbb{R}^D$, even though it acts like a delta function (when it $\star$-multiplies other functions) under integration.
 
\
 
Finally, the same formula expressing the gluing of open spin network vertices, and defining cylindrical functions for closed graphs as special cases of functions associated to a given number of them, can be given in the spin representation.
In fact, the function $\Psi_\gamma$ can be expanded in group representations, using the Peter-Weyl
theorem, as:

\bes \Psi_\gamma(G_{ij}^{ab})=\sum \,\Psi_{\gamma\;\{
m_{ij}^{ab} l_{ij}^{ab}\}}^{J_{ij}}\,\prod_i
\overline{C^{J_{ij}^{ab}\mathcal{I}_i}_{m_{ij}^{ab}}}C^{J_{ij}^{ab}\mathcal{I}_i}_{n_{ij}^{ab}}\prod_{[(ij)(ab)]\in E}
D^{J_{ij}^{ab}}_{s_{ij}^{ab}n_{ij}^{ab}}(G_{ij}^{ab})\,=\,\sum_{\{ J \},\mathcal{I_i}}\tilde{\Psi}_{\gamma\,}^{\{
J_{ij}^{ab} \},\mathcal{I}_i} \prod_{i} C^{J_{ij}^{ab}\mathcal{I}_i}_{n_{ij}^{ab}}\prod_{[(ij)(ab)]\in E}
D^{J_{ij}^{ab}}_{s_{ij}^{ab}n_{ij}^{ab}}(G_{ij}^{ab}) \ees

with representations $J_{ij}^{ab}$ of the
group $G$, representation matrices $D^{J}$, whose
indices refer to the start and end vertex of the edges $[(ij)(ab)]$ to
which the group element $G_{ij}^{ab}$ is attached; $C^{\{
J_{ij}^{ab}\mathcal{I}\}}$ are (normalized) intertwiners for the group
$G$, attached (in pairs) to the vertices, resulting from the
requirement of gauge invariance at the vertices of the graph
$\gamma$, a basis of which is labeled by additional quantum numbers $\mathcal{I}$; these intertwiners contract all indices of both the modes of $\Psi$ and of the representation functions, leaving a gauge invariant function of spin variables only. 

A similar decomposition is of course available for the functions $\varphi$. In terms of these functions, one can then obtain the functions $\Psi$ for closed graphs as:

\bes \Psi_\gamma(G_{ij}^{ab}=g_{i}^a(g_j^{b})^{-1})\,&=&\,\prod_{[(ij)(ab)]}\int d\alpha_{ij}^{ab}
\sum_{\{\vec{J}_{i}\},\{ \vec{m}_{i} \},\mathcal{I}_i}
\,\varphi_{\vec{m}_{i}}^{\vec{J}_{i},\mathcal{I}_i}\,\prod_{i}
\left[ \prod_{j\neq i} D^{J_{i}^a}_{m_{i}^a n_{i}^a}(g_{i}^a
\alpha_{ij}^{ab})\right]\,C^{\vec{J}_j,\mathcal{I}_i}_{\vec{n}_j}\, \nonumber \\
&=&\,\sum_{\{ J_{ij}^{ab}\},\mathcal{I}_i}\Psi_{\gamma}^{\{ J_{ij}^{ab}
\},\mathcal{I}_i}\prod_{i}  C^{J_{ij}^{ab}\mathcal{I}_i}_{n_{ij}^{ab}}\prod_{[(ij)(ab)]\in E}
D^{J_{ij}^{ab}}_{s_{ij}^{ab}n_{ij}^{ab}}(g_{i}^a(g_j^{b})^{-1})
\ees

from which one reads the gluing formula in spin representation:  
\be \Psi_\gamma^{\{ J_{ij}, \mathcal{I}_i \}}=\sum_{\{ m_i^j \}}
\varphi_{\vec{m}_i}^{\vec{J}_i, \mathcal{I}_i}\,\prod_{[(ij)(ab)]\in E}\delta_{J_i^a,J_j^b}\,\delta_{m_i^a,m_j^b}. \ee

\

In the end, the basic point is representation-independent: 

\

any loop quantum gravity state can be seen as a linear combination of (or a special case of) states describing disconnected open spin network vertices, of arbitrary number, with additional conditions enforcing gluing conditions and encoding the connectivity of the graph.

\

Of course, we have considered here only the definition, within $\mathcal{H}_V$, of LQG states associated to graphs with given number $V$ of (d-valent) vertices. In order to deal with graphs with an arbitrary number of vertices, one has to consider the larger Hilbert space 

\be
\mathcal{H}\;=\; \bigoplus_{V=0}^{\infty}\; \mathcal{H}_V \qquad \label{1stHilbert} .
\ee
This larger Hilbert space is what we will consider as the \lq\lq first quantized\rq\rq Hilbert space of reference for the construction for the second quantised theory.

\

What we have shown so far is that we can put in correspondence the whole {\it set} of LQG states in $\mathcal{H}_1$ or $\mathcal{H}_2$ to the set of states in $\mathcal{H}$.

\medskip

We need now to go beyond the level of {\it sets} only, to deal with the vector space structure of the above Hilbert spaces, the LQG one and the new $\mathcal{H}$, i.e. with the scalar product among their states. It is at this level that the relation between the two Hilbert spaces, both similarities and differences, becomes manifest.

First of all, we show that the usual kinematical scalar product between two
cylindrical functions associated to the {\it same} given graph $\gamma$ is expressible in terms of the natural scalar product between
two functions in $\mathcal{H}_V$, modulo a
gauge fixing (which is basically irrelevant in the case of compact group
$G$), due to the gluing conditions. In other words, the scalar product on $\mathcal{H}_\gamma$ (which is also the one used in both $\mathcal{H}_1$ and $\mathcal{H}_2$, once one has fixed the graph) is the one induced by the scalar product in $\mathcal{H}_V$ for the special class of states corresponding to closed graphs.

Indeed, in the case of compact groups, using the above expression for a generic cylindrical
function in terms of the function $\varphi$, one has the simple formula:

\bes \la \Psi_\gamma|\tl{\Psi}_{\gamma}\ra\,&=&\,  \prod_{[(ij)(ab)]\in E} \int_G dG_{ij}^{ab}\,
\overline{\Psi}_\gamma(G_{ij}^{ab})\tilde{\Psi}_\gamma(G_{ij}^{ab}) \,=\, \nonumber \\ &=&
\,\prod_{(ij)}\int dg_i^j dg_j^i \int d\alpha_{ij}
d\tl{\alpha}_{ij}^{ab}\,\overline{\varphi}(...g_i^a
\alpha_{ij}^{ab},..,g_j^b\alpha_{ij}^{ab},...)\,\tl{\varphi}(...,g_i^a\tl{\alpha}_{ij}^{ab},..,g_j^b\tl{\alpha}_{ij}^{ab},...)\,=
\nonumber \\ &=&\,\prod_{[(ij)(ab)]\in E}\int d(g_i^a
(g_j^b)^{-1}) d(g_i^a g_j^b)\,\overline{\varphi}(...g_i^a
(g_j^b)^{-1},...)\,\tl{\varphi}(...,g_i^a
(g_j^b)^{-1},...)
 \ees

where we have used the normalized Haar measure throughout, which makes irrelevant the redundant integration over the
variables $g_i^a g_j^b$, one for each edge of the graph $\gamma$.
That is, we have the standard LQG kinematical scalar product \cite{LQG} induced from the natural scalar product for the functions
$\varphi\in\mathcal{H}_V$ themselves:

\be \la \varphi|\tl{\varphi}\ra\,=\,\prod_{(ij)}\int
dg_i^a \, \overline{\varphi
(\vec{g}_i)}\,\tl{\varphi}(\vec{g}_i)\; .
\ee
Clearly the formula has to be modified in the case of $G$ being a non-compact group (where $V_G$ is infinite), using the necessary gauge fixing.

\

The above shows that we can actually embed $\mathcal{H}_\gamma$ into $\mathcal{H}_V$ (assuming that the graph $\gamma$ has $V$ vertices) faithfully, i.e. preserving the scalar product, and thus the whole $\cup_\gamma \, \mathcal{H}_\gamma$ into the larger $\mathcal{H}\,=\,\bigoplus_V \, \mathcal{H}_V$. 

\

However, it is easy to see that the first quantised Hilbert space $\mathcal{H}$ differs in important ways from both the two possible \lq LQG Hilbert spaces\rq ~$\mathcal{H}_1$ and, more important, $\mathcal{H}_2$. This is at once obvious after noticing that, in defining the Hilbert space $\mathcal{H}$ out of the Hilbert spaces $\mathcal{H}_V \supset \mathcal{H}_\gamma$, we take a direct sum {\it over the number of vertices}, as opposed to the set of graphs, and we {\it do not impose any cylindrical equivalence} (and, thus, in particular, one keeps the zero modes in the Hilbert space).  Indeed, states associated to {\it different} graphs end up being organised very differently in $\mathcal{H}$ compared to the LQG spaces: 

- states associated to graphs with a {\it different number of vertices} are {\it orthogonal}, but

- states associated to {\it different graphs} but with the {\it same number of vertices} are, generically, {\it not orthogonal}.

In a way, the new Hilbert space $\mathcal{H}$ decreases the importance of the exact graph structure, as compared to the LQG Hilbert space(s), while maintaining the number of vertices as an important purely combinatorial attribute of the quantum states. It is this feature that makes it straightforward to give it a Fock space structure. This is what we will do in the next section.

\

The key observation that makes the construction intuitive is that the functions $\varphi(\vec{g}_1,...,\vec{g}_i,...,\vec{g}_V)$ can be understood as \lq many-body\rq ~wave functions for $V$ \lq\lq quanta\rq\rq corresponding to the $V$ open spin network vertices to which the function refers\footnote{To clarify the physical analogy we have in mind, which also motivates the following construction, one can think of the same type of functions $\varphi$ in the case of $G=\mathbb{R}$ and no gauge invariance imposed, so that each single-vertex wavefunction can be seen as a single-particle wave function for a particle on $\mathbb{R}^d$, and the generic function $\varphi$ as a true many-particle system of $V$ particles: $\varphi(x_1,y_1,z_1,...,x_V,y_V,z_V)\equiv\varphi(\vec{x}_1,
..., \vec{x}_V)$. The gluing projection encoded by the $\alpha$
integrations amounts to requiring that the multi-particle wave
function above depends only on the relative position, in some
given direction, of the same particles, for example, requiring
$\varphi(x_1,y_1,z_1,...,x_V,y_V,z_V) = \varphi(x_1 - x_V,....)$, thus defining a special class of symmetric states in the same Hilbert space.}. To further justify the analogy, it suffices to realize that the above functions are exactly the many-particles wave functions for point particles living on the group manifold $G^d$, and having as classical phase space $\left(\mathcal{T}^*G\right)^d$ (which is indeed also the classical phase space of a single spin network vertex or polyhedron).

Accordingly, each state can be conveniently decomposed into products of \lq\lq single-particle\rq\rq (single-vertex) states:

\be
| \varphi \rangle = \sum_{\vec{\chi}_i/ i = 1,..,V} \varphi^{\vec{\chi}_1...\vec{\chi}_V}\,|\vec{\chi}_1\rangle \cdots |\vec{\chi}_V\rangle 
\ee

which in the group representation reads:

\be
\langle g| \varphi \rangle = \sum_{\vec{\chi}_i/ i = 1,..,V} \varphi^{\vec{\chi}_1...\vec{\chi}_V}\,\langle \vec{g}_1|\vec{\chi}_1\rangle \cdots \langle \vec{g}_V|\vec{\chi}_V\rangle 
\ee

and where the complete basis of single-vertex wave functions is given either by the spin network wave functions for individual spin network vertices (labelled by spins and angular momentum projections associated to their $d$ open edges, and intertwiner quantum numbers):

\be
\vec{\chi} = \left( \vec{J}, \vec{m}, \mathcal{I} \right) \;\;\;\rightarrow\;\;\; \psi_{\vec{\chi}}(\vec{g}) =  \langle \vec{g}|\vec{\chi}\rangle = \left[ \prod_{a=1}^{d} D^{J_a}_{m_a n_a}(g_{a})\right]\,C^{J_1...J_d,\mathcal{I}}_{n_1..n_d} \;\;\; ,
\ee

or by a product of non-commutative plane waves constrained by the closure condition for the fluxes:

\be
\vec{\chi} = \left( \vec{X}| \sum_a X_a =0 \right) \;\;\; \rightarrow \;\;\; \psi_{\vec{\chi}}(\vec{g}) =  \langle \vec{g}|\vec{\chi}\rangle = \left[ \prod_{a=1}^{d} E_{g_a}(X_a)  \right] \, \star \, \delta_\star\left( \sum_a X_a\right) \;\;\;\; ,
\ee


which can be understood as the wave function associated with a single polyhedron with fluxes $X_a$ on its $d$ boundary faces and conjugate discrete holonomies $g_a$.
In terms of the labels $\chi_v$, the normalization condition of the functions $\varphi$ reads:

\be \prod_{v=1}^V \int
d\vec{g}_v\,\overline{\varphi}(\vec{g}_1,...,\vec{g}_V)\,\varphi(\vec{g}_1,...,\vec{g}_V)\,=\,\sum_{\{\chi_v\}}\,\overline{\varphi}^{\{\chi_v\}}\,\varphi^{\{\chi_v\}}\,=\,1
\label{norm} \ee using the normalization of single particle wave functions

\be \int
d\vec{g}\,\overline{\psi_{\vec{\chi}}}(\vec{g}_i)\,\psi_{\vec{\chi}'}(\vec{g})\,=\,\delta_{\vec{\chi},\vec{\chi}'}\;.
\ee

\

Notice that the two bases of single-vertex states are related by the composition of Peter-Weyl and non-commutative Fourier transform, which are both unitary transformations \cite{flux}. The (completion of) the linear span of such basis states defines the single-vertex (single-polyhedron) Hilbert space $\mathcal{H}_v$. 
Having realised that this single-vertex Hilbert space is the basic building block of quantum states in $\mathcal{H}$, one also realises immediately that each subspace $\mathcal{H}_V$ can be understood as the tensor product of $V$ copies of the same $\mathcal{H}_v$. With this, the definition of the Fock space representation of $\mathcal{H}$, which we detail in the next section, is intuitive and technically straightforward.

Notice also that one could work instead with the larger Hilbert spaces of non-gauge invariant states $L^2\left( G^E \right)$ and $L^2\left( G^{d \times V} \right)$, not imposing any gauge symmetry at the vertices of spin network graphs, and consider this condition as part of the dynamics. The above construction would proceed identically, with the same final result, but with the basis of single-vertex states now given by the above functions without the contraction of representation function with a $G$-intertwiner, in the spin representation, and thus with single-vertex labels $\vec{\chi} = \left( \vec{J}, \vec{m}, \vec{n}\right)$, or without the $\star$-multiplication by a non-commutative delta function, in the flux representation, thus with single-vertex labels $\vec{\chi} = \left( \vec{X} \right)$. 

\

This single-vertex, or single-polyhedron, Hilbert space is obviously nothing else than the generalization of the Hilbert space for a topological quantum polyhedron that is well studied in the spin foam literature (mainly in the simplicial case), the basic building block of discrete BF theories, for gauge group $G$ \cite{barbieri, baezbarrett, spinnetPolyhedra}. In the case in which $G=SL(2,\mathbb{C})$ (in the Lorentzian case) or $G=Spin(4)$ (in the Riemannian case) and additional geometric conditions (the discrete counterpart of the Plebanski constraint), aka simplicity constraints, are imposed on the same states (either via projectors at the kinematical level or via the dynamics of the theory), the same Hilbert space can be thought of as describing a geometric polyhedron, the basic building block of 4-dimensional spin foam models. If the Immirzi parameter is added to the theory, then the states can be based on $SU(2)$ and the geometricity of the data gets encoded in the embedding of this data into the full $SL(2,\mathbb{C})$ or $Spin(4)$ group, and implemented at the level of the (quantum) dynamics \cite{SF,GFT-holst,GFT-BC}. 

\section{Second quantized formulation of the kinematics of LQG: GFT kinematics}
Having recast generic LQG states in the form of many-body wave functions, and having identified a basis of single-vertex states, the second quantized reformulation of the same quantum theory, based on the Hilbert space $\mathcal{H}$ and an appropriate algebra of quantum geometric operators $\hat{O}$ acting on it, proceeds via standard textbook methods \cite{Fetter}. We now discuss this reformulation, focusing first on the definition of the Fock space of quantum states built out of $\mathcal{H}_v$. Having defined this, we will move on to the definition of the 2nd quantized counterpart of LQG kinematical operators. We will see that the result of this reformulation is exactly the GFT kinematical structure. In the next section, we apply the same reformulation to the dynamics of the theory, showing how a specific GFT dynamics correspond to any given definition of a LQG quantum dynamics, thus completing the general map between the two formalisms.

\

We anticipate here the result of the second quantized reformulation of the Hilbert space of the theory, which is immediate, once the single-vertex (or single-polyhedron) Hilbert space has been identified. 
 
Starting from this single-vertex Hilbert space, one arrives at a pre-Fock space $\tilde{\mathcal{F}}(\mathcal{H}_v) = \bigoplus_{V=0}^\infty\left( \mathcal{H}_v^{(1)} \otimes \mathcal{H}_v^{(2)} \otimes \cdots \otimes \mathcal{H}_v^{(V)}\right)$. This space is then reduced to the space of states satisfying the appropriate statistics, i.e. encoding the symmetry under \lq particle permutations\rq satisfied by the many-particle wave functions. In this case, we recall, we have {\it assumed} that each spin network wave function is symmetric under the automorphism group of the graph on which it is based, which includes vertex relabeling. Such vertex relabeling translates, in their re-interpretation (and re-writing) as many-vertex wavefunctions, into permutation of vertices 

\be
\varphi\left( \vec{g}_1, \vec{g}_2, \dots, \vec{g}_i,\dots, \vec{g}_j,\dots,\vec{g}_V\right) = \varphi\left( \vec{g}_1, \vec{g}_2, \dots, \vec{g}_j,\dots, \vec{g}_i,\dots,\vec{g}_V\right)
\ee

and it is the relevant transformation to determine the quantum statistics. The assumed invariance leads then to {\it choose} the bosonic statistics for the Fock space reformulation of the Hilbert space. We will arrive at:

\be
\mathcal{H} \simeq \mathcal{F}(\mathcal{H}_v) = \bigoplus_{V=0}^\infty \;\text{sym}\left( \mathcal{H}_v^{(1)} \otimes \mathcal{H}_v^{(2)} \otimes \cdots \otimes \mathcal{H}_v^{(V)}\right)
\ee
where only symmetric elements of $\mathcal{H}_v \otimes \mathcal{H}_v \otimes \cdots \otimes \mathcal{H}_v$ are included.

\

The use of bosonic statistics remains, at this point, a choice. The main argument for the symmetry under vertex relabeling it descends from is that such vertex relabeling, and more generally, graph automorphisms, can be understood as the only equivalent of diffeomorphism transformations in the context of abstract, non-embedded graphs that we have chosen for the definition of the 1st quantized LQG theory. A more proper attempt at justifying this assumption would involve discussing in more detail the relation between diffeomorphisms and graph automorphisms, then, if graphs are understood as originally embedded in a smooth manifold \cite{LQGdiffeos}, or between graph automorphisms and simplicial diffeos, as studied in the context of Regge calculus \cite{diffeoRegge}, if one interpret the graphs (and their histories) as dual to a piece-wise flat cellular complex. We will discuss again these issues, briefly, in the concluding section. For now, it should remain clear that the bosonic statistics of our Fock states is an assumption of the construction we are presenting, and that other possibilities should be considered.

\subsection{Change of basis: occupation numbers}

The first preliminary step in the 2nd quantization is to look at the labels (quantum numbers) $\vec{\chi}$ characterizing single-vertex basis states and assume an ordering for their corresponding set, so that one can further label the same set by a single index $a$, running from one to infinity (if indeed, as in this case, the set of possible quantum numbers has infinite elements). This step is basically just needed for notational convenience, and encodes no real physics (or quantum geometry). In the end, the only thing that matters is the ability to identify unambiguously if two vertices have the same quantum numbers or not. Consider for example the usual spin network labels. Concretely, one could first order the open links in each vertex with a label $l$ running from $1$ to $d$, then order the vertex labels as $(j_1,m_1,...,j_l,m_l,...,j_d,m_d, \mathcal{I})$, and finally choose a lexicographic ordering for the whole set of labels, based on this way of writing them. The same type of ordering choice could be made for the flux labels (where the analogy with point particles is even clearer). 

Considering then multi-vertices states, with states characterized by functions $\varphi^{\vec{\chi}_1 \cdots \vec{\chi}_V}$, one can reorder their labels according to the index $a$ introduced above (obviously, the symmetry of the functions under vertex permutations is what keeps the function itself unchanged in such reordering). Each set of single-vertex labels may of course appear several times. Also, because we work out this decomposition for given number of vertices, a normalization condition on the set of labels is implemented. One gets:

\be
\varphi^{\vec{\chi}_1 ... \vec{\chi}_V} = \varphi^{\overbrace{\vec{\chi}^1...\vec{\chi}^1}^{n_1}...\overbrace{\vec{\chi}^a...\vec{\chi}^a}^{n_a} ....} = C\left( n_1,...,n_a,...\right)\hspace{2cm} \sum_{a=1}^\infty n_a = V
\ee

where we have introduced the coefficients $C\left( n_1,...,n_a,...\right)$ which keep track only of the \lq\lq occupation numbers\rq\rq, i.e. the number of times each combination of single-vertex labels appears, and we dropped the reference to the underlying graph, given that, in absence of gluing conditions, there is no need to use the graphical representation of each single-vertex state as a vertex with $d$ open links.
In terms of the new coefficients, the normalization condition is:

\be
\sum_{\{ n_a \}}\, | C\left(n_1,...,n_a,...,n_\infty\right)|^2\sum_{\{\vec{\chi}_i\} / \{n_a\}}\; 1 \, = \, \sum_{\{ n_a \}}\, | C\left(n_1,...,n_a,...\right)|^2\;\frac{V!}{n_1 !...n_\infty !}
\,=\,1
\ee

where $\frac{V!}{n_1 !...n_\infty !} = \sum_{\{\vec{\chi}_i\} / \{n_a\}}\; 1$ is indeed the sum over all the values of
$\{\chi_i\}$ compatible with the number assignment $\{ n_a \}$ for a total finite number $V$ of vertices (given this constraint, clearly most occupation numbers will be zero).

The constraint of finite number of spin network vertices $V$ can and must be removed, if one wants to work with the full set of LQG states, which includes infinitely complex graphs. We do so and allow for $\sum_{a=1}^\infty n_a = \infty$.

\

Moving to a labelling of quantum states by their \lq\lq occupation numbers\rq\rq means moving to a new basis of the Hilbert space of $V$ spin network vertices:

\bes
\varphi\left( \vec{g}_i \right) = \sum_{\vec{\chi}_i} \varphi^{\vec{\chi}_1 ... \vec{\chi}_V}\; \prod_{i\in V} \langle \vec{g}_i|\vec{\chi}_i\rangle = \sum_{\{ n_a\}} \tilde{C}\left(n_1,.., n_a,..\right) \sqrt{\frac{n_1! ... n_\infty !}{V!}}\sum_{\{ \vec{\chi}_i | n_a \}} \prod_{i\in V} \langle \vec{g}_i|\vec{\chi}_i\rangle = \\ = \sum_{\{ n_a\}} \tilde{C}\left(n_1,.., n_a,..\right) \psi_{\{n_a\}}\left( \vec{g}_i \right) = \sum_{\{ n_a\}} \tilde{C}\left(n_1,.., n_a,..\right) \langle g | n_1, ..., n_a, ...\rangle
\label{stateoccupnumbers} \ees

where we indicated the new basis elements by $| n_1, ..., n_a, ...\rangle = | n_1\rangle \cdots | n_a\rangle \cdots | n_\infty\rangle$, their coefficients in the group representation by  $ \langle g | n_1, ..., n_a, ..., n_\infty\rangle=\psi_{\{n_a\}}\left( \vec{g}_i \right)$ and by $\tilde{C}\left(n_1,.., n_a,..\right)$ the normalized states $\varphi$ in the occupation number basis.
These new basis elements encode the re-writing of the Hilbert space $\mathcal{H}$, which, as we have seen, contains all LQG states, as a Fock space $\mathcal{F}(\mathcal{H}_v) = \bigoplus_{V=0}^\infty\left( \mathcal{H}_v^{(1)} \otimes \mathcal{H}_v^{(2)} \otimes \cdots \otimes \mathcal{H}_v^{(V)}\right)$.

\subsection{Fundamental 2nd quantized operators, Fock vacuum and GFT fields}

As customary, we can define a new set of fundamental operators, $\hat{c}_{a} = \hat{c}_{\vec{\chi}}$ and $\hat{c}^\dagger_{a} = \hat{c}^\dagger_{\vec{\chi}}$, satisfying the relations:

\be
\left[ \hat{c}_{\vec{\chi}} , \hat{c}_{\vec{\chi}'}^\dagger \right] = \delta_{\vec{\chi},\vec{\chi}'}\;\;\;\left[ \hat{c}_{\vec{\chi}} , \hat{c}_{\vec{\chi}'} \right] = \left[ \hat{c}_{\vec{\chi}}^\dagger , \hat{c}_{\vec{\chi}'}^\dagger \right] = 0 \hspace{2cm}
\hat{c}_{\vec{\chi}} | n_{\vec{\chi}}\rangle = \sqrt{n_{\vec{\chi}}} | n_{\vec{\chi}} - 1\rangle\;\;\;\;\hat{c}^\dagger_{\vec{\chi}} | n_{\vec{\chi}}\rangle = \sqrt{n_{\vec{\chi}} + 1} | n_{\vec{\chi}} + 1\rangle
\ee

and the {\it occupation number operators} (one for each set of single-vertex labels $\vec{\chi}$):

\be
\hat{N}_{\vec{\chi}}| n_{\vec{\chi}}\rangle = \hat{c}^\dagger_{\vec{\chi}}\hat{c}_{\vec{\chi}} | n_{\vec{\chi}}\rangle = n_{\vec{\chi}}| n_{\vec{\chi}}\rangle \;\;\;\; .
\ee

These fundamental operators {\it create} and {\it annihilate} single-spin network vertices, when acting on the special state given by the {\it Fock vacuum} $| 0 \rangle$, itself identified with the \lq\lq no-space\rq\rq (or \lq\lq emptiest\rq\rq) state in which no degree of freedom of quantum geometry (no spin network at all) is present, that is the one in which all occupation numbers are zero: $|0\rangle = | 0, 0, ...., 0\rangle$ and defined by the property that it is annihilated by all annihilation operators $\hat{c}_{a} = \hat{c}_{\vec{\chi}}$:  $\hat{c}_{\vec{\chi}} | 0 \rangle = 0 \;\;\;\; \forall \vec{\chi}$. 

Starting from the Fock vacuum, these fundamental operators generate all states in the Fock space $\mathcal{F}(\mathcal{H}_v) = \bigoplus_{V=0}^\infty\left( \mathcal{H}_v^{(1)} \otimes \mathcal{H}_v^{(2)} \otimes \cdots \otimes \mathcal{H}_v^{(V)}\right)$, that is all the (linear combinations of the) states $| n_1, ..., n_a, ...,n_\infty\rangle$. 

\

Because,  every state in the usual LQG Hilbert space of closed spin networks can be written as a linear combination or a particular case of multi-vertices states, as we have shown in the previous section, and because, as equation ~\ref{stateoccupnumbers} shows, every state in the corresponding Hilbert space $\mathcal{H}$ can be written in the basis states $| n_1, ..., n_a, ...,n_\infty\rangle$, the isomorphism between this extended Hilbert space and the Fock space $\mathcal{F}(\mathcal{H}_v) = \bigoplus_{V=0}^\infty\; \text{sym}\left( \mathcal{H}_v^{(1)} \otimes \mathcal{H}_v^{(2)} \otimes \cdots \otimes \mathcal{H}_v^{(V)}\right)$ is established. Notice that we have used the notation $\vec{\chi}$ for the index $a$, in order to highlight the meaning of the fundamental operators.

\

From the (linear superposition of) creation and annihilation operators one can define the bosonic {\it field operators}:

\be
\hat{\varphi}(g_1,..,g_d)\equiv\hat{\varphi}(\vec{g})=\sum_{\vec{\chi}}\;\hat{c}_{\vec{\chi}}\; \psi_{\vec{\chi}}(\vec{g}) \hspace{1cm}
\hat{\varphi}^\dagger(g_1,..,g_d)\equiv\hat{\varphi}^\dagger(\vec{g})=\sum_{\vec{\chi}}\; \hat{c}^\dagger_{\vec{\chi}} \;\psi^*_{\vec{\chi}}(\vec{g}) \;\;\; ,
\ee 

satisfying the commutation relations:

\be
\left[ \hat{\varphi}(\vec{g})\,,\,\hat{\varphi}^\dagger(\vec{g}') \right]\,=\, \mathbb{I}_G(g_i, g_i') \hspace{0.5cm} \left[ \hat{\varphi}(\vec{g})\,,\,\hat{\varphi}(\vec{g}') \right] = \left[ \hat{\varphi}^\dagger(\vec{g})\,,\,\hat{\varphi}^\dagger(\vec{g}') \right] \,=\, 0
\ee

where $\mathbb{I}_G(g_i, g_i')\equiv   \int_G dh \prod_{1}^d\delta(g_i h (g_i')^{-1})$ is the identity operator on the space of gauge invariant fields, that is encoding the gauge invariance condition at the spin network vertices.

One recognizes them as the usual fundamental GFT field operators \cite{GFT}, which indeed can be expanded in modes either via Peter-Weyl decomposition, corresponding to the choice $\vec{\chi} = (\vec{J},\vec{m},\mathcal{I})$, or via the non-commutative Fourier transform, corresponding to the choice $\vec{\chi} = (\vec{X} | \sum_{i=1}^d X_i = 0)$ (in which case, the above formula has to be understood as involving a $\star$-multiplication between field modes (creation/annihilation operators) and single-vertex wave functions). It is then clear also in which sense one has a \lq\lq second quantization\rq\rq of the wave function of a single spin network vertex, now turned into an operator. Indeed, it may be convenient to re-name the creation/annihilation operators as field operators in the conjugate space: $\hat{c}_{\vec{\chi}} \equiv \hat{\varphi}_{\vec{\chi}}$ and $\hat{c}^\dagger_{\vec{\chi}} \equiv \hat{\varphi}^\dagger_{\vec{\chi}}$.

\subsection{Kinematical LQG operators in 2nd quantized (GFT) language}

The fundamental operators corresponding to the basic kinematical structure of GFT are thus obtained naturally from the canonical LQG kinematics. 
The same is true for any other kinematical observables of the canonical theory. The general rules for \lq second quantizing\rq ~a canonical many-particle theory give us a simple prescription for turning any kinematical LQG observable into an equivalent 2nd quantized one, i.e. a unique operator acting on the Fock space and expressed as a function of the basic creation/annihilation operators, or of the field operators \cite{Fetter}.  

\

To start with, consider a 1st quantized \lq 2-body\rq operator $\hat{O}_2$, i.e. an operator acting on a single spin network vertex and resulting in another single spin network vertex, thus not changing the combinatorial structure of the initial simple graph but only the labels (group or spins or fluxes, etc) associated to it. The matrix elements of such operator in a complete basis of $\mathcal{H}$, given by the product of single-vertex states, are $\langle \vec{\chi}| \hat{O}_2 | \vec{\chi}'\rangle$. We have considered only the irreducible components of the operator, thus its matrix elements on single vertex states in $\mathcal{H}$, the others being obtained from it, in the sense that the full operator acting on generic $N$-vertex states will be given by the sum of operators like the above each acting on a single-vertex only.

From the above matrix elements, one deduce immediately the 2nd quantized version of the same operator as:

\be
\widehat{\mathcal{O}_2}\;\;\; \rightarrow \;\;\;\langle \vec{\chi} | \widehat{\mathcal{O}_2} | \vec{\chi}' \rangle = \mathcal{O}_2\left( \vec{\chi}, \vec{\chi}'\right) \hspace{1cm}  \rightarrow \;\;\;\;\;\widehat{\mathcal{O}_2}\left(\varphi\right) = \sum_{\vec{\chi},\vec{\chi}'} \mathcal{O}_2\left( \vec{\chi}, \vec{\chi}'\right) \hat{c}^\dagger_{\vec{\chi}} c_{\vec{\chi}'}=\int  d\vec{g}\,d\vec{g}'\;\mathcal{O}_2\left( \vec{g}, \vec{g}'\right) \widehat{\varphi}^\dagger(\vec{g}) \widehat{\varphi}(\vec{g}')
\ee 

The construction works in exactly the same fashion for more general operators, acting on several spin network vertices, possibly changing the combinatorial structure of the underlying graph as well, for example creating/annihilating spin network vertices. For \lq $(n,m)$-operators\rq ~$O_{n,m}$, acting on $m$ vertices and resulting in $n$ vertices, one gets:

\bes
\widehat{\mathcal{O}_{n,m}}\;\;\; &\rightarrow& \;\;\;\langle \vec{\chi}_1, ....,\vec{\chi}_m  | \widehat{\mathcal{O}_{n,m}} | \vec{\chi}'_1 , ... , \vec{\chi}'_n \rangle = \mathcal{O}_{n,m}\left( \vec{\chi}_1,...,\vec{\chi}_m, \vec{\chi}'_1,...,\vec{\chi}'_n\right) \hspace{1cm}  \rightarrow \nonumber \\ &\rightarrow& \;\;\;\;\;\widehat{\mathcal{O}_{n,m}}\left(\hat{\varphi},\hat{\varphi}^\dagger\right) = \int  d\vec{g}_1...d\vec{g}_m\,d\vec{g}'_1...d\vec{g}'_n\; \widehat{\varphi}^\dagger(\vec{g}_1)...\widehat{\varphi}^\dagger(\vec{g}_m)\mathcal{O}_{n,m}\left( \vec{g}_1,...,\vec{g}_m, \vec{g}'_1,...,\vec{g}'_n\right) \widehat{\varphi}(\vec{g}'_1)...\widehat{\varphi}(\vec{g}'_n)\;\;\;\;
\ees 

The general recipe is then clear: take the matrix elements of the relevant operator, in 1st quantization, in a basis of (the appropriate subspaces of) $\mathcal{H}$; these are complex numbers, function of the set of relevant variables for the $n+m$ spin network vertices involved; take the appropriate convolutions of such functions with creation and annihilation operators depending on the same variables, according to which spin network vertices are acted upon by the operator and which spin network vertices result from the same action, to obtain the 2nd quantized counterpart of the initial LQG operator; this will be an operator obtained as a linear combination of polynomials of creation and annihilation operators, or of GFT field operators. Notice that it is easier to write the 2nd quantized version for extensive observables and 1-body operators, rather than intensive ones.

In general, the action of LQG operators will be non-trivial only on particular combinations of spin network vertices states; for example, gauge invariant (with respect to the action of $G$) operators will have matrix elements involving specific gauge invariant linear combinations of spin network vertex states, those corresponding to spin network vertices glued along open links to form closed graphs. We will give an example of this type of operators in the following. 
\

This concludes the proof of correspondence between canonical LQG states and operators and GFT states and operators, at the kinematical level. In the next section, we lift this equivalence to the dynamical level.

\subsection{Aside: orientation dependence in LQG, spin foams and GFT from a second quantized perspective}
Before we do so, let us digress briefly to discuss the issue of orientation dependence in spin foam models (and thus, in canonical LQG and in GFT). The issue is: does the spin foam dynamics (or equivalently, the corresponding lattice gravity path integral) depend on the orientation of the underlying cellular complex? if not, how can its definition be modified to incorporate such dependence? At a more general level: is the physical scalar product of the canonical theory supposed to be obtained, in a covariant language, through an orientation-independent spin foam (GFT) transition amplitude? do other types of transition amplitudes play a role in quantum gravity (as they do in ordinary relativistic quantum mechanics and quantum field theory)? These questions have been tackled from a variety of perspectives, over the years, in the spin foam literature, starting with \cite{LivineOriti,OritiFeynman}, where the issue of orientation dependence was also related to the issue of incorporating a pre-geometric analogue of causality in spin foam models, and more recently as well \cite{SF-orientation}. Also, the issue has reappeared recently in the context of loop quantum cosmology \cite{LQC-orientation}.

\

The 2nd quantized reformulation of canonical LQG and the corresponding map to a GFT formalism can suggest a new strategy for addressing the issue. The suggestion stems from the fact that, in QFT, the analogue of the \lq\lq orientation dependence\rq\rq of the quantum amplitudes is the need to distinguish, in the formulation of the theory, particles from antiparticles, that is the decomposition of the single-particle Hilbert space into two sectors corresponding to positive and negative energies, in turn related to the implementation of discrete symmetries (parity and time reversal). The orientation reversal operation at the discrete level of Feynman diagrams translates in fact into going from particles to anti-particles, and an \lq\lq orientation dependent\rq\rq transition amplitude is one that distinguishes one from the other. That a similar relation holds in the quantum gravity domain has been already suggested (see the cited references, as well as \cite{thesis}), with particle/antiparticle dichotomy replaced by the space/antispace one in terms of spacetime orientation reversal. The 2nd quantized formulation of LQG seems the right language to make this analogy solid.

\

We have seen that the straightforward 2nd quantization of LQG kinematics leads to a definition of quantum fields that is very close to the standard non-relativistic one used in condensed matter theory, and that is fully compatible with the kinematical scalar product of the canonical theory. In turn, this can be seen as coming directly from the definition of the Hilbert space of a single tetrahedron (more generally, polyhedron). It is known that this contains states corresponding to tetrahedra of opposite orientation, but with the same geometry, related to one another by a parity operator \cite{barbieri,baezbarrett}. One could try to split this Hilbert space into two sectors corresponding to the two opposite orientations, and identify them with the quantum geometry analogue of particle/antiparticle sectors. This also means defining \lq\lq relativistic-like\rq\rq GFT fields 

\be
\hat{\varphi}(g_1,..,g_d)\equiv\hat{\varphi}(\vec{g})=\sum_{\vec{\chi}_+}\left[ \hat{c}_{\vec{\chi}_+} \psi_{\vec{\chi}_+}(\vec{g})+\hat{d}_{\vec{\chi}_+}^\dagger \psi^*_{\vec{\chi}_+}(\vec{g})\right] \qquad
\hat{\varphi}^\dagger(g_1,..,g_d)\equiv\hat{\varphi}^\dagger(\vec{g})=\sum_{\vec{\chi}_+}\left[ \hat{c}^\dagger_{\vec{\chi}_+} \psi^*_{\vec{\chi}_+}(\vec{g})+\hat{d}_{\vec{\chi}_+} \psi_{\vec{\chi}_+}(\vec{g})\right]
\ee

where we have split the modes into positive and negative sectors $\vec{\chi}_{\pm} = \left( \vec{j}, \pm \vec{m}, I \right) \;\;\;\;\;\text{or}\;\;\;\;\;
\vec{\chi}_{\pm} = \left( \pm\vec{X} \right)$. This splitting can be properly defined, for example, by generalizing the set of labels to include the representations of a discrete group, e.g. generalizing the group $G$ to its (semi-)direct product with $\mathbb{Z}_2$, acting as a discrete parity transformation. For example, one could think of  generalizing from the rotation group or Lorentz group to the Pin group of the appropriate dimension \cite{thesis}. We leave this possibility for future work.

Obviously, this new definition requires a modification of the commutation relations, but also, more importantly, a new definition of the kinematical scalar product, that would not be the usual LQG one. 
A dynamics that knows the difference between the two sectors, constructed in more direct analogy with relativistic QFT, would then lead to orientation dependent spin foam amplitudes (thus new spin foam models). In turn, these should be expected to correspond, in the flux representation, to discrete gravity path integrals which do not symmetrize over the two volume orientations or the two sectors of solutions of Plebanski (simplicity) constraints.
We leave any such construction, based on the alternative definitions above, to future work. Most importantly, any future work on these issues should clarify if there exists, at the LQG, spin foam or GFT level, some symmetry requirement that, analogously to Lorentz symmetry in relativistic QFT, forces us to distinguish spacetime orientations, and how this can be related, in Lorentzian models, to any notion of quantum geometric causality.

\section{Second quantized formulation of the dynamics of LQG: GFT dynamics}
The general correspondence between 1st quantized (LQG) and 2nd quantized (GFT) operators, presented in the previous section, allows us to clarify the correspondence between canonical LQG and GFT in a straightforward manner, along the same lines, and without passing through the spin foam formalism, whose relation with canonical LQG will rather be defined and understood {\it via the GFT framework}, that is via the Feynman perturbative expansion of the GFT dynamics corresponding to any given choice of LQG dynamics. 

\subsection{General correspondence}
The canonical quantum LQG dynamics \cite{LQG, QSD,thomasMaster} is encoded in the action, on spin network states, thus on states in $\mathcal{H}$, of a specific $(n+m)$-body operator, the Hamiltonian constraint operator\footnote{We are neglecting here, of course, the subtle issues concerning the fact that the Hamiltonian constraint operator has to act on the dual space of distributions over spin network states, and various other functional analytic subtleties \cite{LQG}. These are certainly essential for a proper definition of the operator and of its action, but not so much for the main point we are focusing on here, that is the correspondence between canonical LQG and GFT quantum dynamics.}. Indeed, the quantum dynamics of the theory is encoded in an equation of the type:

\be
\widehat{H}\,| \Psi \rangle \,=\, 0\;\;\;,
\ee
where $\widehat{H}$ is the Hamiltonian constraint operator\footnote{One could also consider imposing the finite version of the same equation, that is $\widehat{e^H} |\Psi \rangle = | \Psi \rangle$, which is equivalent to the infinitesimal one, under very general conditions on the operator $\widehat{H}$ itself, or imposition through a master constraint equation.}. Equivalently, the same dynamics can be encoded in a \lq projection \rq operator onto physical states, that is onto solutions of the above Hamiltonian constraint equation \cite{projector}. In this case, the relevant equation encoding the full quantum dynamics is of the form:

\be
\widehat{P} \, | \Psi \rangle \, = \, | \Psi \rangle
\ee
that is, physical states are those left invariant by the projector operator $\widehat{P}$, which itself may be constructed from the Hamiltonian constraint operator or directly (this will be the case in the explicit example we will discuss later in this section). We will focus on this way of encoding the dynamics, in the following, also because it is the one implemented in all known spin foam/GFT models, as we will discuss.

According to the general rules for translating a 1st quantized operator on 2nd quantized language, outlined in the previous section, the relevant quantities to be used are the matrix elements of the dynamical operator, here $\widehat{P}$, between basis states in $\mathcal{H}$. In general, such operator will decompose into 2-body, 3-body,..., $(n+m)$-body operators, i.e. into operators whose action involves 2,3,...,$(n+m)$ spin network vertices (in turn distinguishable into $(n,m)$ operators, thus with non-zero matrix elements in all these sectors of $\mathcal{H}$. This decomposition may well involve an infinite number of components, unless restricted by symmetry considerations or other special conditions on the projector operator. Beside symmetry and other consistency conditions, the physical question to be addressed in specific models is which terms in this decomposition are actually relevant and whether higher order terms can be reduced to lower order ones, if not in a strict mathematical sense, at least in the sense of a justified physical approximation. 

We weight this decomposition of the general operator $\widehat{P}$ with suitable coupling constants, and write:

\be
\widehat{P} \Psi \rangle = | \Psi \rangle  \rightarrow \left[\lambda_2\widehat{P}_2 + \lambda_{3}\widehat{P}_3 + ....\right] | \Psi \rangle = | \Psi \rangle
\ee  

and for each $(n,m)$-body component of $\widehat{P}$ we consider the matrix elements:

\be
\langle \vec{\chi}_1, ....,\vec{\chi}_m  | \widehat{P_{n+m}} | \vec{\chi}'_1 , ... , \vec{\chi}'_n \rangle = P_{n,m}\left( \vec{\chi}_1,...,\vec{\chi}_m, \vec{\chi}'_1,...,\vec{\chi}'_n\right) 
\ee

in a complete basis of products of single-vertex states. 

Using the general map \cite{Fetter}, we turn then the projector operator into a 2nd quantized one, acting on the Fock space constructed in the previous section, and constructed from linear combination of polynomial of creation/annihilation (GFT field) operators, obtaining the 2nd quantized form of the LQG dynamical equation:

\be
\widehat{F} | \Psi \rangle \equiv \sum_{n,m }^\infty \lambda_{n,m} \;\left[ \sum_{\{ \vec{\chi},\vec{\chi}'\} }\hat{c}_{\vec{\chi}_1}^\dagger ... \hat{c}_{\vec{\chi}_m}^\dagger \; P_{n,m}\left( \vec{\chi}_1,...,\vec{\chi}_m, \vec{\chi}'_1,...,\vec{\chi}'_n\right) \; \hat{c}_{\vec{\chi}_1'} ... \hat{c}_{\vec{\chi}_n'}\; -\;\sum_{\vec{\chi} }\hat{c}_{\vec{\chi}}^\dagger \hat{c}_{\vec{\chi}}\,\right] | \Psi \rangle = \; 0  \;\;\;\;
\ee 

where now the quantum state $| \Psi\rangle$ is understood as a state in $\mathcal{F}(\mathcal{H}_v) = \bigoplus_{V=0}^\infty\; \text{sym}\left( \mathcal{H}_v^{(1)} \otimes \mathcal{H}_v^{(2)} \otimes \cdots \otimes \mathcal{H}_v^{(V)}\right)$.

The same equation can be written in the group representation, giving:

\be
\hspace{-0.5cm} \widehat{F} | \Psi \rangle \equiv \sum_{n,m /}^\infty \lambda_{n,m} \;\left[\int  d\vec{g}_1...d\vec{g}_m\,d\vec{g}'_1...d\vec{g}'_n\; \widehat{\varphi}^\dagger(\vec{g}_1)...\widehat{\varphi}^\dagger(\vec{g}_m)P_{n,m}\left( \vec{g}_1,...,\vec{g}_m, \vec{g}'_1,...,\vec{g}'_n\right) \widehat{\varphi}(\vec{g}'_1)...\widehat{\varphi}(\vec{g}'_n) \; -\;\int d\vec{g}\; \widehat{\varphi}^\dagger(\vec{g})\widehat{\varphi}(\vec{g}) \right] | \Psi \rangle = \;0\;\;\; \label{QDynamicsGroup}.
\ee

This is the operator form of the GFT quantum dynamical equations, which can be alternatively encoded in the Schwinger-Dyson equations for n-point functions\footnote{In usual relativistic QFT on (curved) spacetime \cite{localQFT,Strocchi}, the fundamental field operators have to be understood as operator-valued {\it distributions}, due to UV divergences which make the notion of field at a point not well-defined. A proper definition of the interaction operator, thus, requires some care. Ultimately, this is a necessary consequence of causality, locality and relativistic symmetries. In non-relativistic quantum field theory, such as those used in condensed matter theory, on the other hand, such requirements are not imposed and fields do not have to be treated as distributions, avoiding some of the functional analytic troubles. The case of GFTs is open, because the appropriate replacement for the notion of locality and the symmetries of different models still require more understanding. Moreover, the straightforward 2nd quantization of LQG, as presented here, leads to a formulation that is closer to non-relativistic field theories in condensed matter systems than it is to relativistic QFTs of particle physics. For all these reasons, many of the functional analytic aspects of GFTs as field theories have still to be clarified, and we should be wary of importing conventional field theoretic wisdom into this spacetime-free context. On the other hand, we do expect to face all the difficulties of conventional field theories that are directly due to the infinite number of degrees of freedom involved. For this reason, unless the number of GFT quanta is kept fixed or suitable cut-offs (both UV and IR) are maintained, some of our formulae have to be understood as formal. The very use of the kinematical Fock space to define the theory is known to be troublesome for interacting quantum field theories, and Haag's theorem tells us that we should not expect solutions to the quantum dynamics to be defined as elements on this Fock space. The work in \cite{GFTcondensate} can be seen as a first step towards understanding the structure of more realistic vacua. Many of these issues (e.g. definition of quantum dynamic operator and of its space of solutions, removal of cut-offs, identification of new physical representations, non-perturbative definition of the theory) have a strict canonical LQG analog \cite{LQG} and the field-theoretic GFT framework \cite{GFT,tenGFT} offers new tools to study them, in particular renormalization and constructive techniques \cite{VincentBook}}. 

\

We have chosen the normal ordering for creation/annihilation operators, as customary in many-body physics \cite{Fetter}. Other choices of operator ordering can be considered, resulting in quantum corrections to the above. These quantum corrections can be absorbed into a redefinition of the kernels $P_{n,m}$ or, if their functional form is left invariant, into a \lq renormalization\rq of the coupling constants $\lambda_{n,m}$ (this is one more reason why we do not insist on a specific value for these coupling constants). It is important to keep in mind the presence of such quantum corrections, in order to understand better (the subtleties of) the correspondence with the GFT quantum dynamics.

\

Let us discuss how the above 2nd quantized quantum dynamical operator equation leads to the identification of the corresponding GFT model in terms of a classical action, encoding the same dynamics. Here is where the present level of understanding of the LQG quantum dynamics, of its spin foam counterpart, and of the GFT encoding of the same, forces us to proceed in a more tentative, certainly formal and heuristic manner. Both the quantities we will work with on the canonical side, and their counterparts on the GFT side will be lacking a rigorous mathematical definition. Still, we will see that our tentative conclusions will be at least compatible with present knowledge. Moreover, the formal correspondence we put in place will also clarify exactly how progress on the precise definition of any of the two sides of the correspondence will help a more rigorous definition of the other.

\

A first choice we make is to phrase the question in a quantum statistical language, that is, to try to answer it in the context of a definition of a quantum statistical partition function for the theory\footnote{We are well aware of the many difficulties in formulating quantum statistical mechanics in a generally covariant context and, even more, in a background independent context, to to be applied to quantum spacetime itself (see for example \cite{CarloStatMech}). Our tentative steps can be seen as a suggestion to start a similar programme in the general LQG/GFT context, taking advantage of the 2nd quantized formulation and of the analogy with quantum many-body systems.}.

The starting point is of course the operator equation \ref{QDynamicsGroup}. Knowing this, once would like to define a partition function $Z$ for the quantum LQG theory, in turn defined in terms of a statistical density operator $\hat{\rho}$. 
The first obvious choice would be to define an analogue of the {\it microcanonical ensemble}: $\hat{\rho}_m = \frac{\delta( \widehat{F} )}{Z_m}$ with:

\be
Z_m \, =\,  \sum_{s} \langle s | \;\delta( \widehat{F})  | s\rangle \qquad ,
\ee

where $s$ denotes an arbitrary complete basis of states in the Hilbert (Fock) space of the quantum theory. 

This choice would mean that one defines a (quantum statistical) dynamics in which only states solving the dynamical equation \ref{QDynamicsGroup} contribute. It would be the obvious option in a strict canonical continuum quantum gravity theory, in which physical states of quantum spacetime are only those solving the canonical constraint operator. However, it is not the only possible option in a generalized theory that on the one hand is {\it defined} in a more abstract, algebraic and discrete context (where the very notion of diffeomorphisms and other continuous symmetries is more ambiguous) and on the other hand includes in its covariant \lq histories\rq, i.e. dynamical processes, spacetimes that would not admit (in the continuum) a canonical quantization procedure. Of course, an example of the latter is a path integral prescription for the dynamics which includes topology change for space and/or spacetime.

One may expect from general arguments, and we will indeed confirm it in the following, that the GFT dynamics corresponds to a quantum LQG dynamics of this generalized type. This generalized quantum dynamics in quantum statistical terms amounts to a choice of  a density operator of the {\it canonical} type\footnote{Obviously the term \lq canonical\rq ~is used here in a different sense than in \lq canonical quantization\rq, that is in the sense of ensembles.}: $\hat{\rho}_c = \frac{e^{-\,\widehat{F}}}{Z_c}$ with:

\be
Z_c \, = \, \sum_{s} \langle s | e^{-\,\widehat{F}}  | s\rangle \qquad ,
\ee

where we have put any \lq\lq inverse temperature\rq\rq ~constant $\beta = 1$, for simplicity. This choice corresponds to allowing a non-zero weight to quantum states not solving the constraint equation \ref{QDynamicsGroup}, but at the same time weighting more such solutions compared to generic states (and so that only such solutions are obtained in the limit $\beta\rightarrow\infty$, where one reproduces the microcanonical ensemble). 

In the same spirit, one can weight differently quantum states with more or less \lq particles\rq, that is many or few spin network vertices, and move to a {\it grandcanonical} ensemble, defined by the density operator $\hat{\rho}_c = \frac{e^{-\,\left( \widehat{F} \, -\, \mu \widehat{N}\right)}}{Z_g}$, with

\be
Z_g \, = \,  \sum_{s} \langle s | e^{-\, \left(\widehat{F} \, -\, \mu \widehat{N}\right)}  | s\rangle \qquad ,
\ee

where the sign of the \lq chemical potential\rq $\mu$ determines whether quantum states with many or few spin network vertices are favored. In the following, we work in this general context. We will show that, indeed, this is the choice that corresponds to existing group field theories, clarifying in the process what one should expect when studying the correspondence between the corresponding spin foam amplitudes and canonical LQG.

\

Now we want to rewrite the above partition function in GFT terms, that is as a GFT path integral. The way we proceed is the field theory analogue\footnote{With the obvious functional analytic issues, which we neglect in the following, as our goal is to show the correspondence between LQG and GFT, not to solve the issue of defining in rigorous mathematical terms either the first or the second.} of the standard procedure of constructing the coherent state path integral in quantum mechanics \cite{chaichian}. Indeed, the 2nd quantized formulation of \ref{QDynamicsGroup} is already set up to make this procedure the most convenient one.

We introduce then a 2nd quantized basis of eigenstates of the annihilation operator, that is the GFT quantum field operator, $| \varphi \rangle  = e^{ \sum_{\vec{\chi}} \varphi_{\vec{\chi}} \widehat{c}_{\vec{\chi}}^\dagger} |0 \rangle = e^{ \int d\vec{g}\, \varphi(\vec{g}) \widehat{\varphi}(\vec{g})^\dagger} |0 \rangle$ satisfying:

\bes
&&\widehat{c_{\vec{\chi}}} | \varphi \rangle = \varphi_{\vec{\chi}} \, | \varphi \rangle  \qquad \langle \varphi | \widehat{c_{\vec{\chi}}}^\dagger  = \overline{\varphi_{\vec{\chi}}} \, \langle \varphi | \\ \text{or equivalently}\qquad &&
\widehat{\varphi(\vec{g})} | \varphi \rangle = \varphi(\vec{g}) \, | \varphi \rangle  \qquad \langle \varphi | \widehat{\varphi}^\dagger(\vec{g})  = \overline{\varphi(\vec{g})} \, \langle \varphi | \\
&&\mathbb{I} = \int \mathcal{D}\varphi \mathcal{D}\overline{\varphi}\, e^{-\, |\varphi|^2}\, | \varphi \rangle \langle \varphi | \qquad |\varphi|^2 \equiv \int d\vec{g} \,\overline{\varphi}(\vec{g})\,\varphi(\vec{g}) \, =\, \sum_{\vec{\chi}}\, \overline{\varphi_{\vec{\chi}} }\,\varphi_{\vec{\chi}} \qquad
\ees
 
 The functions $\varphi$ and $\overline{\varphi}$ can be understood as the classical GFT fields, as we are going to see\footnote{These are exactly the simple GFT condensate states that have been given a cosmological interpretation in \cite{GFTcondensate}}. The measure of integration $\mathcal{D}\varphi\mathcal{D}\overline{\varphi}$ is the (formally defined) functional measure for fields on $G^d$ entering the GFT path integral. In terms of this basis of states (and using the bosonic statistics of the GFT states), the partition function reads:
 
 \be
Z_g = \sum_{s} \langle s | e^{-\, \left(\widehat{F} \, -\, \mu \widehat{N}\right)}  | s\rangle \, = \,   \int \mathcal{D}\varphi \mathcal{D}\overline{\varphi}\, e^{-\, |\varphi|^2}\, \langle \varphi |\, e^{-\, \left(\widehat{F} \, -\, \mu \widehat{N}\right)} \,  | \varphi \rangle \qquad .
\ee

It is clear that we have here the GFT path integral with a quantum amplitude

\be
e^{-\, |\varphi|^2}\, \langle \varphi |\, e^{-\, \left(\widehat{F} \, -\, \mu \widehat{N}\right)} \,  | \varphi \rangle \, \equiv\, e^{- \, S_{eff}}
\ee

that can be expressed in terms of an effective action $S_{eff}(\varphi,\overline{\varphi})$. The task of relating the partition function of the quantum LQG theory, defined by some specific operator $\widehat{F}$ to the GFT path integral of  specific models turns into the task of understanding the form of the effective action $S_{eff}$. 

For generic operators $\widehat{F}$, this is a very difficult task \cite{chaichian}, of course. The general strategy for solving it, however, is clear: one should  a) expand the exponential operator function of $\widehat{F}$ in power series of polynomials of creation/annihilation operators, and then b) normal order them, in such a way that their expectation value in the coherent state basis gives a functional of GFT fields according to the general formula:

\bes
\langle \varphi |\, \left[\int  d\vec{g}_1...d\vec{g}_m\,d\vec{g}'_1...d\vec{g}'_n\; \widehat{\varphi}^\dagger(\vec{g}_1)...\widehat{\varphi}^\dagger(\vec{g}_m)\mathcal{O}\left( \vec{g}_1,...,\vec{g}_m, \vec{g}'_1,...,\vec{g}'_n\right) \widehat{\varphi}(\vec{g}'_1)...\widehat{\varphi}(\vec{g}'_n)\right]  \,  | \varphi \rangle = \nonumber \\ = \, \int  d\vec{g}_1...d\vec{g}_m\,d\vec{g}'_1...d\vec{g}'_n\; \overline{\varphi}(\vec{g}_1)...\overline{\varphi}(\vec{g}_m)\mathcal{O}\left( \vec{g}_1,...,\vec{g}_m, \vec{g}'_1,...,\vec{g}'_n\right) \varphi(\vec{g}'_1)...\varphi(\vec{g}'_n) \qquad ;
\ees 

finally, c) one should reassemble the terms in such an expansion and recognize the power series expansion of the function $e^{-\,S_{eff}}$, which can then be read out. 

While the tedious, straightforward calculation gives a result that depends, of course, on the specific operator $\widehat{F}$ one started from, a couple of properties are general: 

\

i) the effective action $S_{eff}$ is obtained from the {\it classical action} $S_0$ by adding quantum corrections, with this classical action given by:

\be
S_{eff}\left(\varphi,\overline{\varphi}\right) \, =\, S_0\left(\varphi,\overline{\varphi}\right)\, +\, \mathcal{O}(\hbar)\, =\, \frac{\langle \varphi | \widehat{F} | \varphi \rangle}{\langle \varphi | \varphi \rangle} \, + \, \mathcal{O}(\hbar) \qquad ;
\ee 

where it is immediate to see that the coupling constant playing the role of the chemical potential in the quantum statistical definition of the partition function becomes a mass term in the classical GFT action, rescaling the term coming from the identity operator $m^2 = 1 + \mu$.

\

ii) the quantum corrections, arising from the normal ordering of terms in the power series expansion, amount to new interaction kernels to be added to the classical ones or, in the simpler cases, to a redefinition of the coupling constants for the same interactions.

\

Notice, then, that the same type of quantum corrections would have appeared, as we pointed out earlier, if we had chosen a different operator ordering in the very definition of the initial dynamical operator $\widehat{F}$. Therefore, rather than focusing on a specific quantum operator $\widehat{F}$ and try to derive the exact effective action $S_{eff}$, to be then used in the GFT path integral corresponding to it, one could as well start from a classical action $S_0$ and interpret the corresponding path integral as coming from some quantum corrected operator $\widehat{F}'$ (plus chemical potential term). 

Moreover, this path integral, just like the quantum statistical partition function of LQG theory, would have anyway to be properly defined, and this definition amounts exactly to control the quantum corrections corresponding to the possible interaction processes of LQG/GFT states (spin networks), i.e. the spin foams, redefine the interaction kernels to take these quantum corrections into account, renormalize the theory \cite{GFTrenorm}, and, finally, try to give non-perturbative meaning to the GFT perturbative (spin foam) expansion. In the end, what is essential is that one can identify the classical GFT action $S_{0}\left(\varphi,\overline{\varphi}\right)$ corresponding to a given LQG dynamical operator $\widehat{F}$, define the corresponding quantum theory, formally at first (e.g. through its perturbative expansion in Feynman diagrams, i.e. spin foams) and then more rigorously, and, possibly, use the field theory formalism to extract effective continuum physics in interesting regimes. 
It is also for the above reasons that we do not focus too much, at this stage, on the precise values of the coupling constant of the theory, appearing either in the quantum operator equations, or in the GFT action.

\

For a given operator equation \ref{QDynamicsGroup}, then, the corresponding classical (and bare) GFT action is of the form:

\bes
S\left( \varphi, \varphi^\dagger\right)\; =\; m^2 \int d\vec{g}\; \varphi^\dagger(\vec{g})\,\varphi(\vec{g})\; - \hspace{12cm} \nonumber \\ -\;\sum_{n,m / n+m = 2}^\infty \lambda_{n+m} \;\left[\int  d\vec{g}_1...d\vec{g}_m\,d\vec{g}'_1...d\vec{g}'_n\; \varphi^\dagger(\vec{g}_1)...\varphi^\dagger(\vec{g}_m)\;V_{n+m}\left( \vec{g}_1,...,\vec{g}_m, \vec{g}'_1,...,\vec{g}'_n\right) \varphi(\vec{g}'_1)...\varphi(\vec{g}'_n)\right] \quad \\ V_{n+m}\left( \vec{g}_1,...,\vec{g}_m, \vec{g}'_1,...,\vec{g}'_n\right) = P_{n+m}\left( \vec{g}_1,...,\vec{g}_m, \vec{g}'_1,...,\vec{g}'_n\right) \nonumber
\ees

where we have highlighted the fact that the GFT interaction kernels are nothing else than the matrix elements of the canonical projector operator in the basis of (products of) single-vertex states, and we have included a non-trivial mass term to incorporate the chemical potential.

\

At this point, our formal correspondence teaches us already an important lesson: 

the spin foam vertex amplitudes, which are nothing else than such GFT interaction kernels, have to be understood as encoding the matrix elements of the projector operator of the canonical LQG theory, and not directly those of the Hamiltonian constraint operator. 

\

This has been suggested and studied before in the literature \cite{AlesciNouiSardelli, thomasantonia} and it is not a new insight per se; what is new, here, is only the justification of this fact from a natural correspondence between canonical LQG dynamics and GFT one, in turn explaining the correspondence with the relative spin foam model. It is clear, from the same general correspondence, that this would not be the case, had we started from the Hamiltonian constraint equation directly instead of the projector equation, at the 1st quantized level. However, as we said, the latter seems to be the choice encoded in all known spin foam/GFT models, one example of which we will discuss in the next subsection. 

Notice also that the use of the projector equation, involving the identity operator, leads directly to the usual trivial kinetic term of many GFT models. It can be interpreted, from a QFT point of view, as a mass term with the value of the mass having been absorbed in the other coupling constants. The $\widehat{P}_2$ component of the projector operator, involving the action on single-vertices, and not changing the graph structure of the spin network states it acts on, would instead contribute further possibly non-trivial kinetic terms, for example those of the laplacian type also being used in the GFT literature \cite{generalizedGFT,ValentinJoseph}, especially in the context of GFT renormalization \cite{GFTrenorm}.

\

Finally, we point out another general consequence of the established correspondence: 

the quantum dynamics of the canonical LQG theory has to be looked for, in the GFT and spin foam reformulation, in the full set of Schwinger-Dyson equations for its n-point functions and in associated Ward identities (following from specific symmetries of the dynamics) \cite{GFT,laurentGFT}. In turn, these can obviously be studied perturbatively in terms of the spin foam expansion of the n-point functions, but this, as in standard QFT, is expected to be a good approximation of the full dynamics only for processes involving very few quantum geometric degrees of freedom of the kinematical type, i.e. only for physical situations in which the physical vacuum of interest is well approximated by the perturbative Fock vacuum \cite{GFTfluid,VincentTensor,tenGFT}.

\subsection{Example: 3d LQG and generalised Boulatov GFT}
The previous construction is very general. It is important, though, to provide some explicit example of it, not only to make it concrete, but also to show that it does match what we do know about both the canonical LQG and the GFT side of the story. 

\

The best context to do so it 3d Riemannian quantum gravity without cosmological constant, in its BF formulation. In this case we have, on the one hand, a good understanding of the canonical quantization of the theory using LQG techniques, and in particular the projector operator encoding the quantum dynamics is known \cite{AlexKarim}. On the other hand, we have a more or less established GFT model describing the same physical system, the Boulatov model, whose corresponding spin foam model is the well-known Ponzano-Regge model \cite{PonzanoRegge}. This has been studied extensively not only from the spin foam point of view, but also in several of its QFT aspects: symmetries \cite{LaurentDiffeos,GFTdiffeos}, renormalization \cite{matteovalentin,GFTrenorm}, summability \cite{LaurentBorel}, critical behaviour \cite{melonicPhase}. It is important then to show that these two sides are related nicely by our general 2nd quantization procedure. Moreover, the correspondence between the canonical LQG theory and the Ponzano-Regge model has been elucidated nicely by Noui and Perez \cite{AlexKarim}, who showed that the canonical scalar product between generic spin network states (i.e. the matrix elements of the projector operator between generic spin network states) can be computed via a sum over a particular class of complexes (containing no \lq\lq bubbles\rq\rq and thus no divergences) weighted by the Ponzano-Regge spin foam amplitudes. In order to make contact with these results, we restrict ourselves to the simplicial context, thus we consider $d=3$, that is trivalent spin network only.

\

The first thing to notice is that the dynamical projector operator of the theory encodes {\it both the spatial diffeomorphism and the Hamiltonian constraints}. The classical constraint imposes a local flatness condition $F^i(\omega(x)) = 0$ at each point of the spatial manifold ($F$ is the 2-form curvature of the 1-form connection $\omega$,  both) valued in the $\mathfrak{su}(2)$ Lie algebra. The LQG quantization leads to replacing connection and curvature Lie algebra elements with group-valued parallel transports and holonomies, turned into acting on spin network states. The projector operator $\widehat{P}$ imposing the flatness condition on quantum states is defined through its matrix elements on arbitrary spin network states $\Psi_\gamma$ and $\Psi'_{\gamma'}$, based on the graphs $\gamma$ and $\gamma'$, which take the form \cite{AlexKarim}:

\be
\langle \Psi_\gamma | \widehat{P} | \Psi'_{\gamma'} \rangle = \langle\Psi_\gamma | \prod_{f \in \bar{\gamma} \supset \gamma, \gamma'} \delta\left(H_f\right) | \Psi'_{\gamma'}\rangle
\ee  

where: $\bar{\gamma}$ is a closed graph containing both graphs $\gamma$ and $\gamma'$, involving in particular the gluing of their open links, if any; $f$ are the {\it independent} closed loops on $\tilde{\gamma}$; $H_f$ are the holonomies around these independent closed loops\footnote{The fact that we consider the projector operator as a matricial one represents a slight abuse of notation, as discussed in \cite{AlesciNouiSardelli} and \cite{AlexKarim}, as the operator should really be thought of as a map from the space of cylindrical functions on $\bar{\gamma}$ to the complex numbers.}. The fact that only closed graphs can appear is due to the gauge invariance of the projector operator.

In order to apply the 2nd quantization translation of this operator, and extract the corresponding terms in the GFT action, following the procedure we presented above, we need the matrix elements of the same operator between $n+m$-vertices states.

Given this generic expression, we realize that the projector operator will in general have non-zero matrix elements between any {\it even} number of spin network vertices only, due to the combination of choice of valence for the same vertices and of gauge invariance of the operator. That is, the projector operator decomposes into a combination of $2-,4-, 6-,...,$body operators. At each order, moreover, different terms will contribute, corresponding to the different closed graphs that can be formed by $m+n$ open spin network vertices.
Also, a priori we have to consider as different cases the ones corresponding to $n$ \lq\lq initial\rq\rq spin network vertices and $m$ \lq\lq final\rq\rq spin network vertices, the opposite case of $m$ initial and $n$ final vertices, and all the various combinations with $n+m$ vertices in total. We may label them $P_{0,2}, P_{1,1}, P_{2,0}$, and $P_{0,3}, P_{1,2}, P_{2,1}, P_{3,0}$, and so on. They give rise to different 2nd quantized operators, since the data corresponding to initial spin network vertices are convoluted with annihilation operators, and those for final vertices with creation operators. In this specific case, however, one can verify that $P_{n,m}=P_{n+m}$, so all these different 2nd quantized operators (and their corresponding GFT interaction terms) have the same kernel. Only a closer look at symmetry properties and other requirements (e.g. some generalized notion of locality) to be imposed on the dynamics would allow to restrict the number of terms that do contribute to the quantum dynamics, while simple considerations of hermiticity of the projector operator (and reality of the GFT action) require that a term of the type $P_{m,n}$ is always accompanied by a term of the type $P_{n,m}$.

\

We now move on to consider some of the corresponding matrix elements $P_{m+n}$, which, as we have seen, become then the interaction kernels $\mathcal{V}_{m+n}$ of the GFT action. We assume, for simplicity, that open spin network links outgoing from the same vertex cannot glue to one another, forming \lq tadpole-like\rq closed graphs (the only motivation for this assumption, at this stage, is that these graphs would correspond to degenerate polyhedra). However, we do not impose any additional restriction.

The calculation is simpler in the group representation, and in the same representation we present the results.

\

\begin{figure}[h]
\includegraphics[scale=0.3]{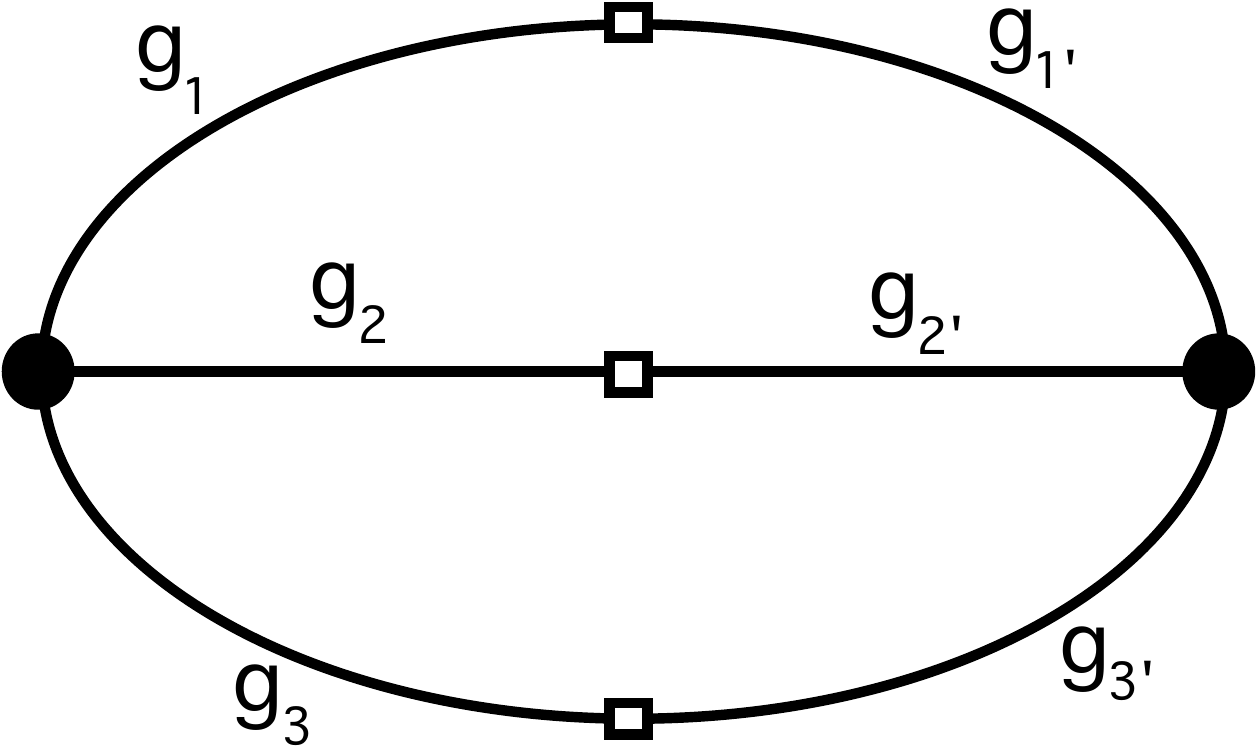}
\caption{The unique closed graph formed by two 3-valent spin network vertices: the \lq\lq melon\rq\rq or \lq\lq dipole\rq\rq}
\label{melon}
\end{figure}

There is a single graph $\bar{\gamma}$ to consider, formed by two spin network vertices only, the 3d (super-)melon or dipole graph of figure ~\ref{melon}. 
The corresponding matrix element is $P_2 = \delta\left( G_1 G_2^{-1}\right)\delta\left( G_2G_3^{-1}\right)$, with $G_i = g_i (g_i')^{-1}$, and where $g_i$ group elements label the half links.

\begin{figure}[h]
\includegraphics[scale=0.3]{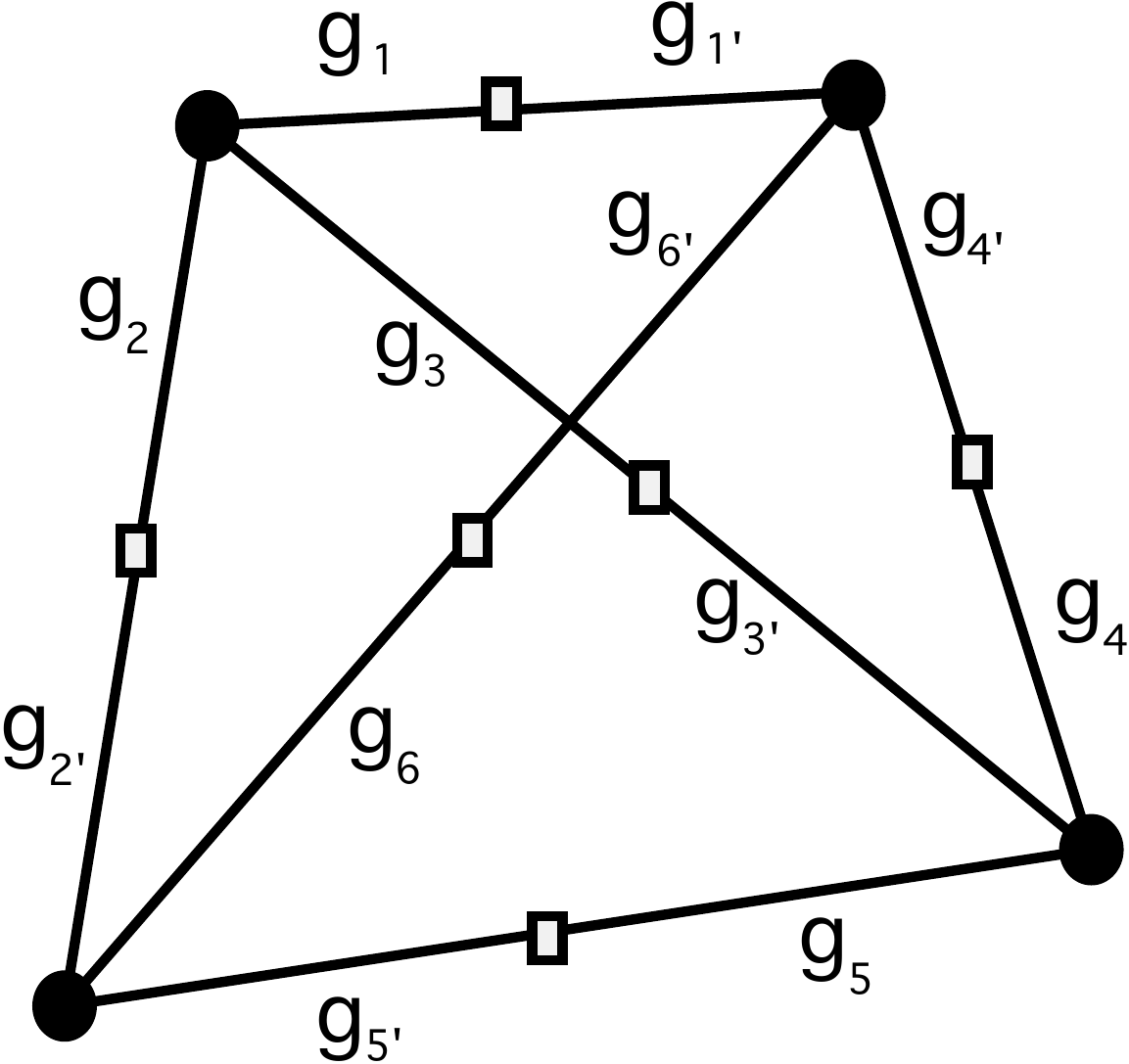}
\caption{The tetrahedral graph formed by four 3-valent spin network vertices}
\label{tetra}
\end{figure}

At order 4, there are two graphs to consider. One is the tetrahedral graph represented in figure ~\ref{tetra} , with the same notation for group elements labeling its half links.

The matrix element is $P_4^{tet} = \delta\left( G_3 G_5 G_2^{-1}\right)\delta\left( G_2 G_6 G_1^{-1}\right) \delta\left( G_4 G_6^{-1} G_5^{-1}\right)$.

\begin{figure}[h]
\includegraphics[scale=0.3]{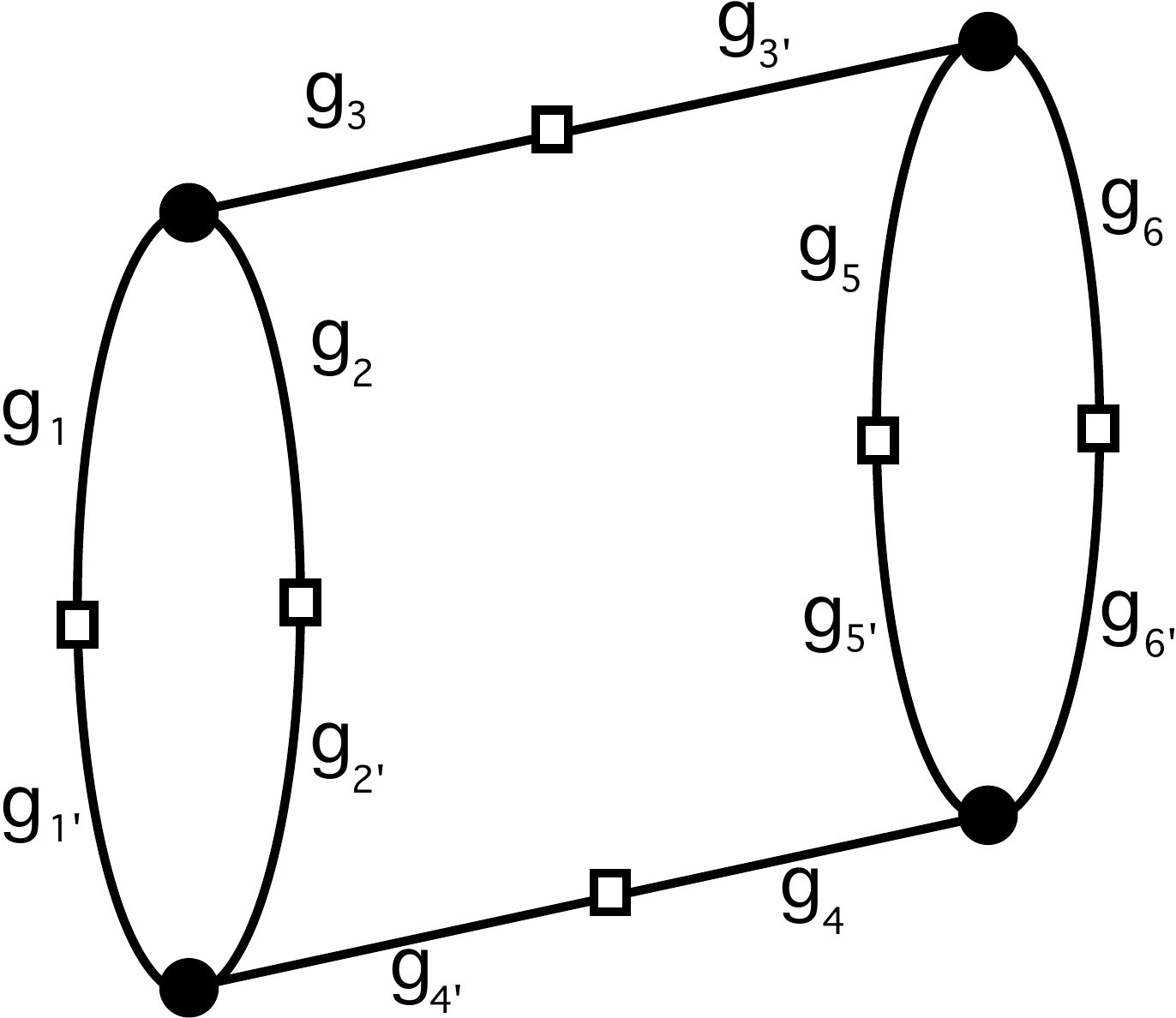}
\caption{The \lq\lq pillow graph\rq\rq formed by four 3-valent spin network vertices}
\label{pillow}
\end{figure}

The other graph is the one in figure ~\ref{pillow}, with corresponding matrix element $P_4^{pil} = \delta\left( G_2 G_1\right)\delta\left( G_6 G_5\right) \delta\left( G_3 G_5 G_4 G_2\right)$.

\begin{figure}[h]
\hspace{-0.7cm}
\begin{minipage}[b]{3cm} 
\includegraphics[scale=0.15]{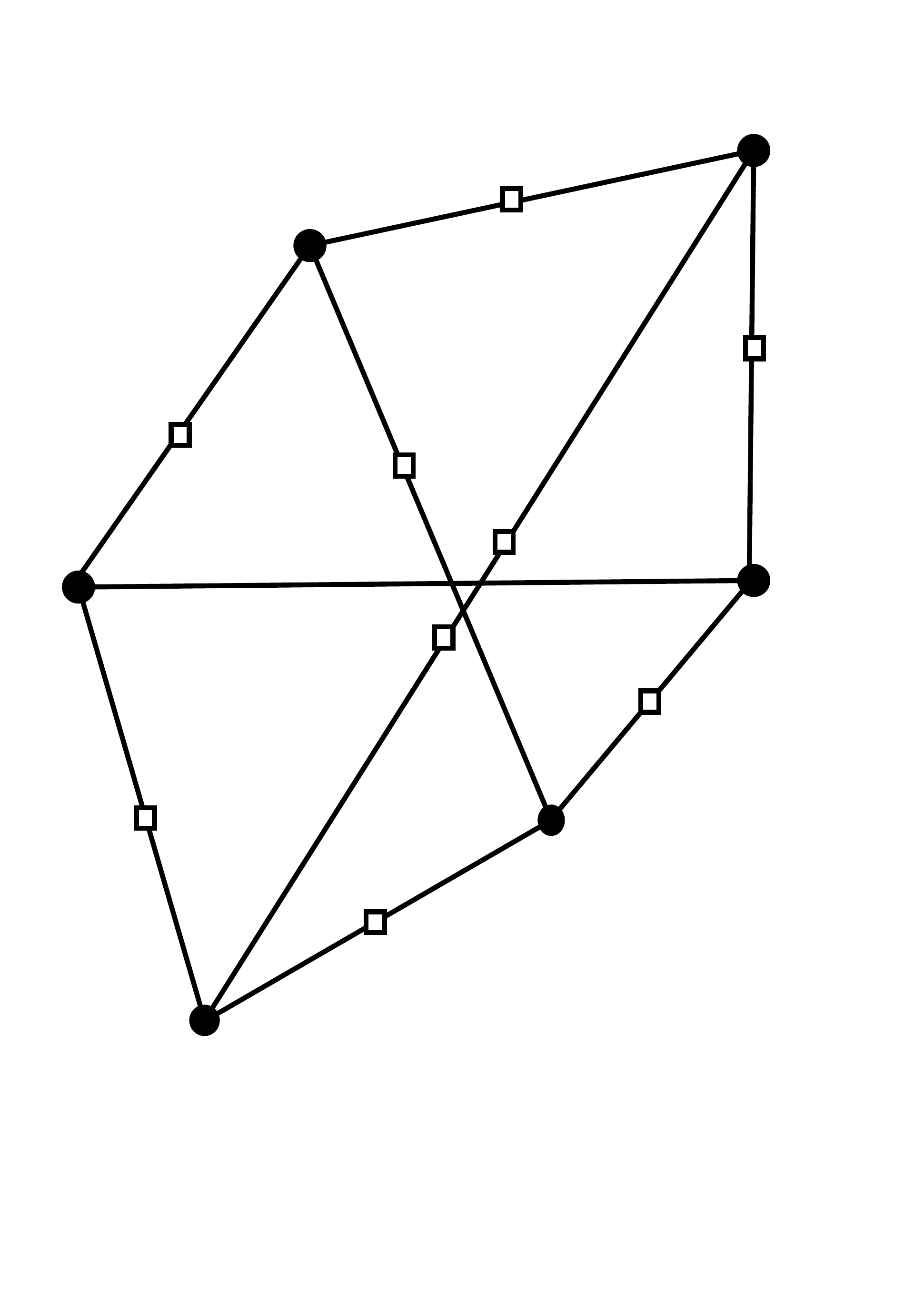} 
\end{minipage}
\qquad
\hspace{-0.7cm}
\begin{minipage}[b]{3cm} 
\includegraphics[scale=0.15]{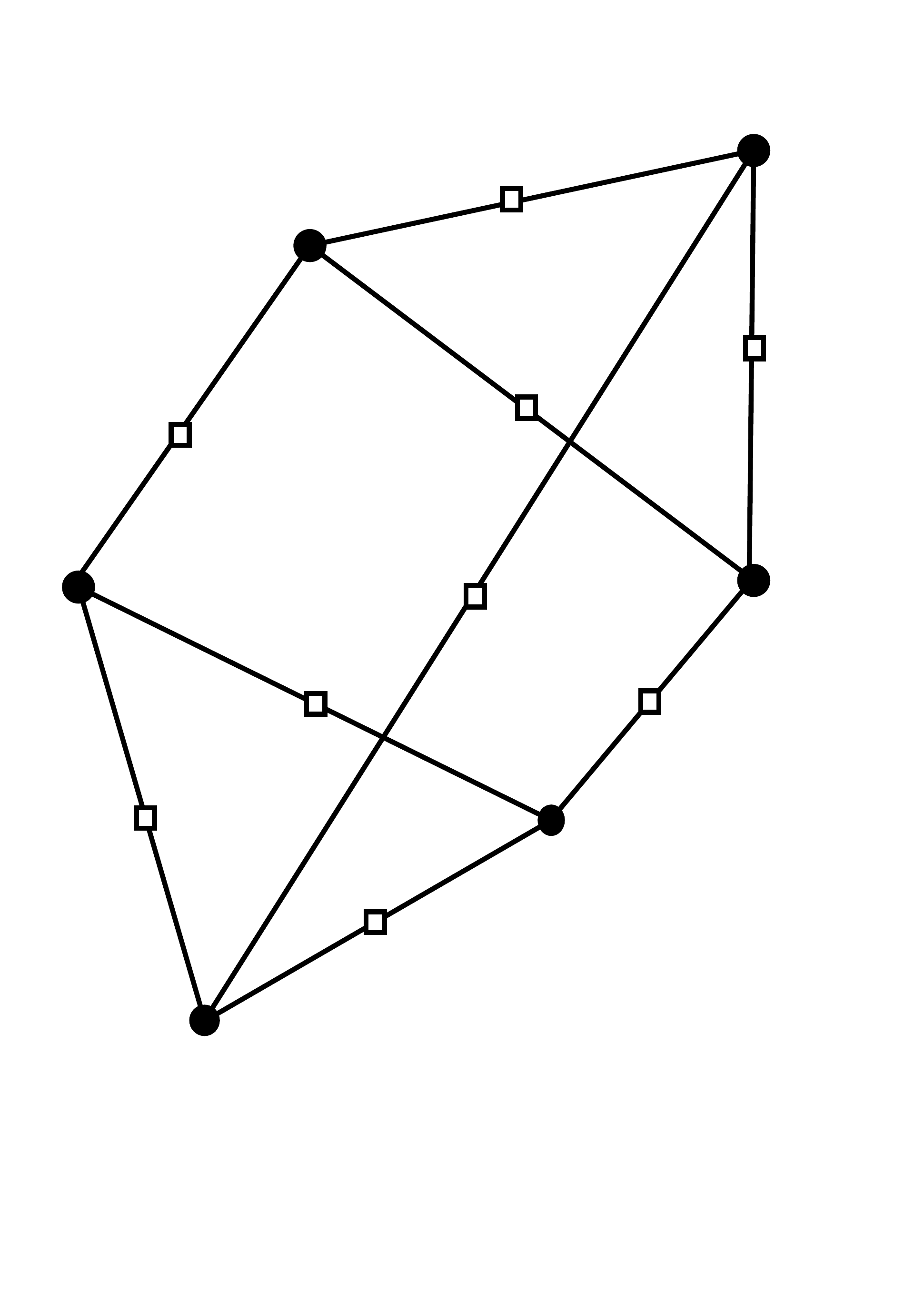} 
\end{minipage}
\qquad
\begin{minipage}[b]{3cm}
\includegraphics[scale=0.15]{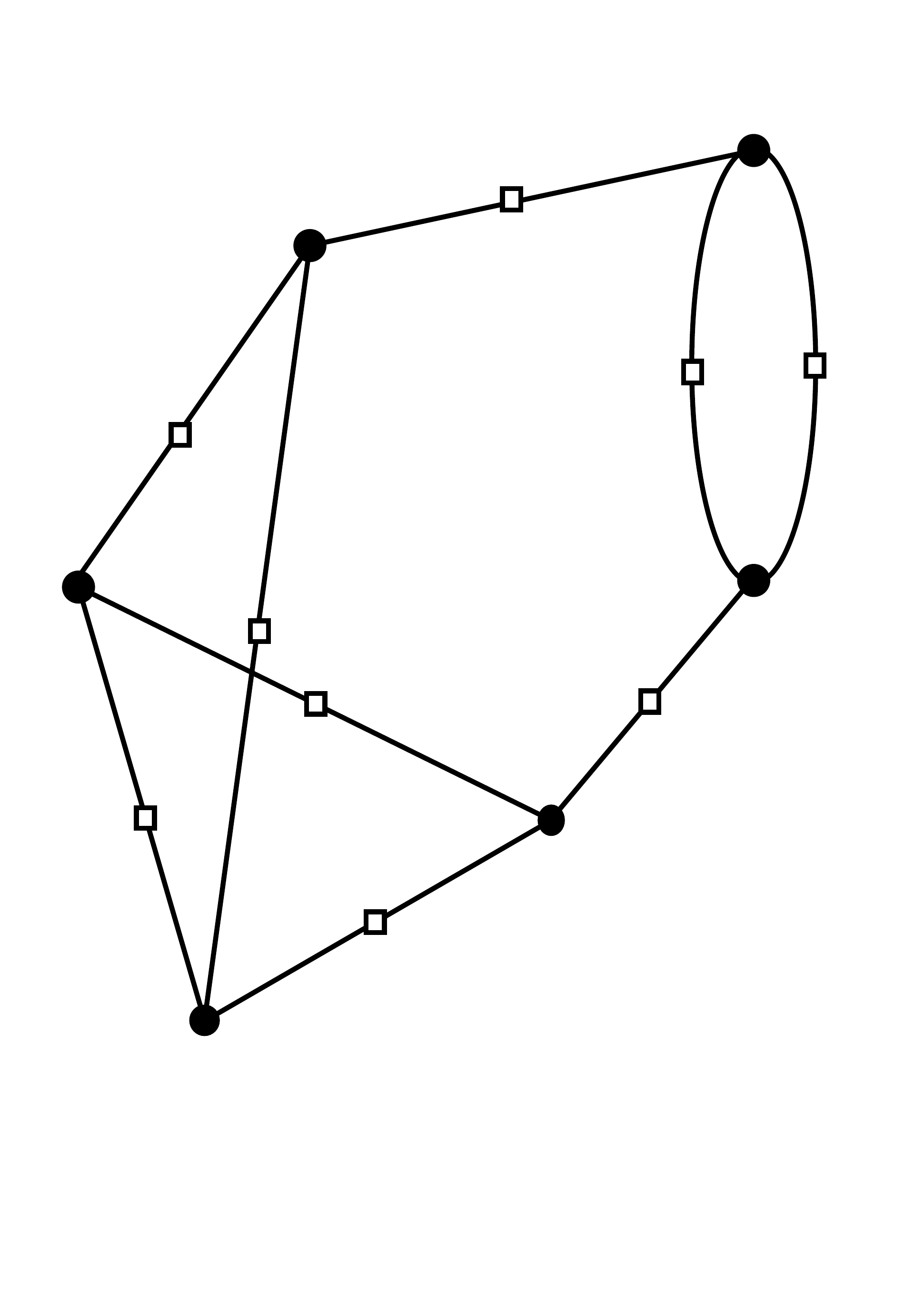} 
\end{minipage}
\qquad
\begin{minipage}[b]{3cm}
\includegraphics[scale=0.15]{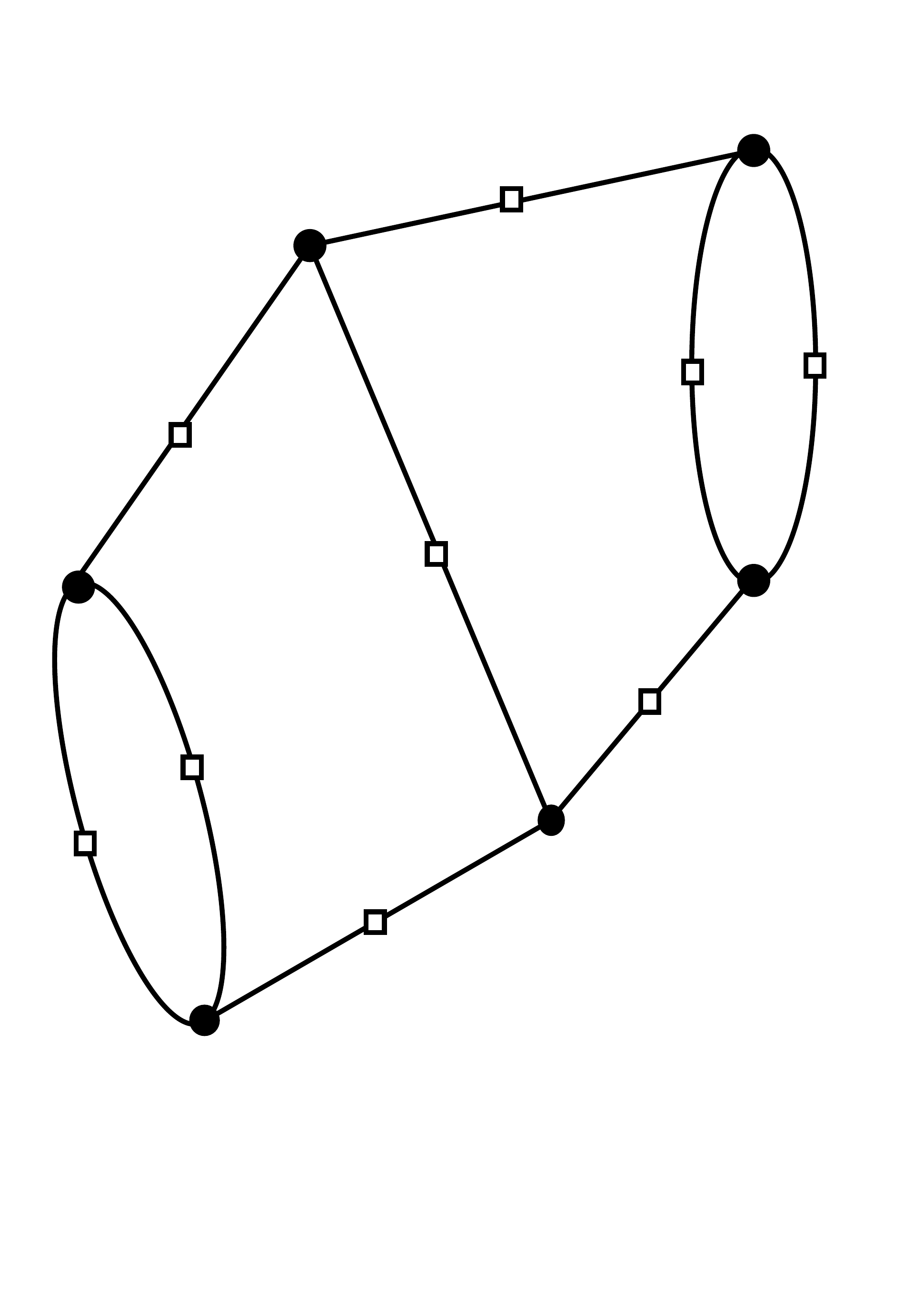}
\end{minipage}
\quad
\begin{minipage}[b]{3cm}
\includegraphics[scale=0.15]{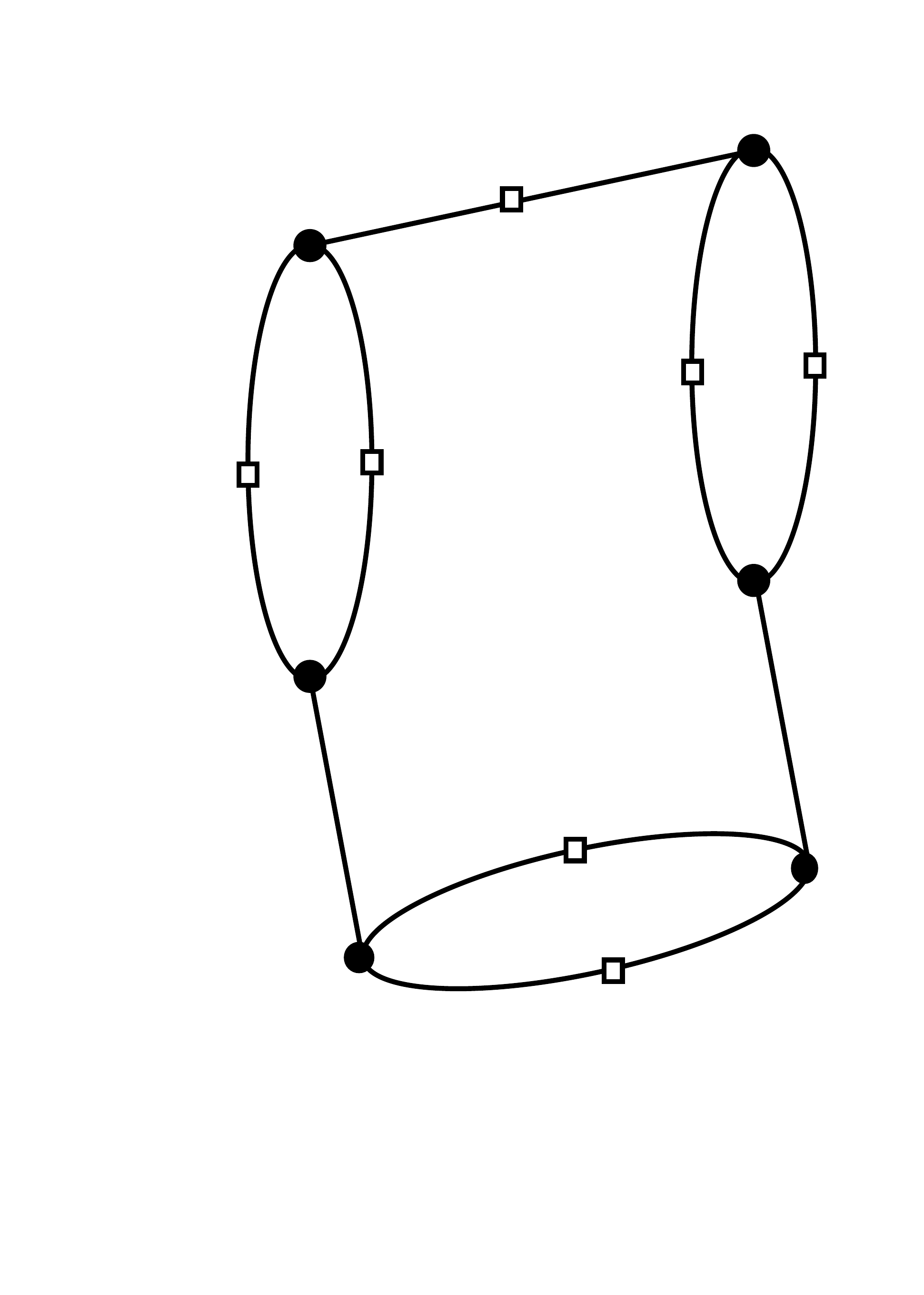}
\end{minipage}
\caption{Closed graphs obtained from six 3-valent spin network vertices, contributing to the dynamical operator}
\label{order6}
\end{figure}

Obviously, more graphs contribute at order 6, 8, etc, and correspond to more contributions to the spin network and GFT quantum dynamics. For example, the five graphs contributing at order $m+n=6$ are shown in figure ~\ref{order6}.

Having the matrix elements of the projector operator, one can construct the 2nd quantized dynamical operator, with one contribution from each matrix elements above (distinguishing the various terms with the same kernel $P_{n+m}$), weighted by some coupling constant, and from there, as we have shown in the general case, the terms in the corresponding GFT action. Here, we report directly these latter terms, showing only one of the one having the same kernel $P_{m,n}=P_{m+n}$, the one corresponding to the case $P_{0,n}$.
 
In performing the following manipulations, corresponding to simple change of variables under group integration, one uses only the gauge invariance conditions for the GFT fields (open spin network vertices).

It is immediate to see that the single term at order 2 gives the identity operator, thus contributes only a rescaling of the constant in front of the identity operator in the dynamical equation, or of the \lq mass\rq ~term in the GFT action.

For the tetrahedral term at order 4, we get:

\bes
\int \left[ dg_i dg_{i'}\right]\varphi_{123} \;\varphi_{3'45} \; \varphi_{5'2'6}\; \varphi_{6'4'1'} \; \delta\left( G_3 G_5 G_2^{-1}\right)\,\delta\left( G_2 G_6 G_1^{-1} \right)\,\delta\left( G_4 G_6^{-1}G_5^{-1} \right) \;= \;...\; = \;\;\;\;\;\; \hspace{2cm} \nonumber \\ =\;\int \left[ dg_i dg_{i'}\right]\varphi_{123} \;\varphi_{3'45}\; \varphi_{5'2'6}\;\varphi_{6'4'1'}\; \delta\left( g_1 g_{1'}^{-1} \right)\,\delta\left( g_2 g_{2'}^{-1} \right)\,\delta\left( g_3 g_{3'}^{-1} \right)\,\delta\left( g_4 g_{4'}^{-1} \right)\,\delta\left( g_5 g_{5'}^{-1} \right)\,\delta\left( g_6 g_{6'}^{-1} \right)\;\;\;\; \\ G_i = g_i g_{i'}^{-1} \;\;\;\;\;\; \varphi_{ijk} = \varphi\left( g_i, g_j, g_k \right) \nonumber\;\;\; .
\ees

One recognizes immediately the usual tetrahedral interaction of the standard Boulatov GFT model \cite{boulatov,GFT}. Indeed, the fact that the Boulatov term encodes the projector operator of the canonical theory is well-known and understood from a variety of perspectives \cite{PonzanoRegge}. For example, it was clarified in the analysis of diffeomorphisms at the GFT level \cite{GFTdiffeos} which in turn suggested to rewrite the Bouatov model first in the flux representation, with Lie algebra elements $x_i$ associated to the edges of the triangle corresponding to the GFT field, and then move to Lie algebra variables associated to the vertices of the same triangles. In terms of these variables, in fact, the discrete diffeomorphisms corresponding to vertex translations act naturally. Indeed, these vertex variables $v_i$ are exactly the Lagrange multipliers (discretized lapse and shift) appearing in the rewriting of the component of the projector operator associated to each independent loop on the spin network graph in the flux representation, for example: 

\be
\delta\left( G_3 G_5 G_2^{-1}\right) = \int_{\mathbb{R}^3 \simeq \mathfrak{su}(2)} dv\, E_{G_3 G_5 G_2^{-1}}(v)\,=\, \int_{\mathbb{R}^3 \simeq \mathfrak{su}(2)} dv\, e^{\frac{i}{2}Tr(v G_3 G_5 G_2^{-1})}
\ee

where $E_g(x)$ are the non-commutative plane waves on which the flux representation \cite{GFTdiffeos} relies, and where we recognize the very definition of the (discretized form of the) canonical projector operator. A more indirect connection can be established considering the recursion relations satisfied by the 6j-symbol \cite{recursion} (the Boulatov interaction term), and which were also shown to arise from the diffeomorphism symmetry of the GFT action \cite{GFTdiffeos}.

The other term of order 4 gives instead:

\bes
\int \left[ dg_i dg_{i'}\right]\varphi_{123} \;\varphi_{3'56} \; \varphi_{5'46'}\; \varphi_{4'2'1'} \;\delta\left( G_2 G_1\right)\delta\left( G_6 G_5\right) \delta\left( G_3 G_5 G_4 G_2\right) \;= \;...\; = \;\;\;\;\;\; \hspace{2cm} \nonumber \\ =\;\int \left[ dg_i dg_{i'}\right]\varphi_{123} \;\varphi_{3'56}\; \varphi_{5'46'}\;\varphi_{4'2'1'}\; \delta\left( g_1 g_{1'}^{-1} \right)\,\delta\left( g_2 g_{2'}^{-1} \right)\,\delta\left( g_3 g_{3'}^{-1} \right)\,\delta\left( g_4 g_{4'}^{-1} \right)\,\delta\left( g_5 g_{5'}^{-1} \right)\,\delta\left( g_6 g_{6'}^{-1} \right)\;\;\;\;  .
\ees
 
 This term is also well known in the GFT literature as the \lq\lq pillow term\rq\rq, that has been studied in the context of GFT renormalization \cite{GFTrenorm}, and that has been shown to lead to Borel summability of the model, when added (with appropriate sign of the coupling constant) to the standard Boulatov term \cite{LaurentBorel}.  
 
One can repeat the same simple check for the other, higher order terms contributing to the dynamics, and verify that, for gauge invariant fields, the flatness condition imposed by the projector operator translates into simple identifications between group elements associated to the spin network vertices (thus to the GFT fields). One can also verify that, thanks to the simple nature of kinetic and interaction kernels, the quantum corrections that relate the classical GFT action formed by the above terms (including all orders) with the quantum projector operator, and its full (generalized) quantum statistical partition function, can all be absorbed in the coupling constants of the same classical action (after regularization), i.e. they do not generate new types of interaction kernels.

\

We have thus verified that the canonical LQG dynamics is indeed encoded in a generalized Boulatov GFT, in which (an a priori infinite number of) extra interaction terms are added to the standard tetrahedral one. These additional terms lead, in the perturbative expansion of the theory, to a generalized class of complexes obtained as Feynman diagrams of the theory, on top of the simplicial ones, but all weighted by the same spin foam amplitudes, the Ponzano-Regge model, all corresponding to the discrete BF path integral in the non-commutative flux representation. It should also be noted, though, that the spin foam expansion of the GFT dynamics obtained in this way directly from the canonical theory would differ from the one obtained in \cite{AlexKarim}. The latter, as said, only sums over complexes without 3-cells (and therefore with finite quantum amplitudes), and seems to corresponds to a gauge-fixed version of the more general one. How this gauge-fixing is obtained, from the 2nd quantized perspective, in the procedure used in \cite{AlexKarim} (which treats somehow different contributions to the projector operator in succession, rather than in parallel, and inserts a spin network decomposition of the identity between each of them), or, equivalently, to which regime of the full GFT expansion it corresponds, needs to be clarified and interpreted in field theory terms. 

\

Anyway, this is a concrete example of the general correspondence between canonical LQG and covariant GFT, via 2nd quantization methods, which also establish a direct link between canonical LQG and spin foam models, via their GFT formulation. It should be clear, then, that the same construction can be performed for 4d gravity models, and the analysis of \cite{AlesciNouiSardelli} could be a good starting point, together with the more recent results of \cite{thomasantonia}. We note that also in all these 4d models, because the GFT action (mean value of the quantum dynamical LQG operator in the coherent state basis) is obtained by constraining the action of the GFT model for BF theory, the vertex kernel (i.e. the spin foam vertex amplitude) has to be as encoding the matrix elements of the canonical projector operator.

\section{Comments and outlook}
We have seen that canonical LQG can be reformulated in 2nd quantized language to define a group field theory, and that this general correspondence works both at the kinematical and dynamical level: to any definition of a LQG theory (a specific choice of kinematical Hilbert space and observables and of quantum dynamics) corresponds a specific GFT (with Fock space, 2nd quantized observables and quantum dynamics), and viceversa.

Beside the pedagogic value of clarifying the link between the GFT formalism and canonical LQG, the established correspondence is also a clear, definite link between canonical LQG and covariant spin foams, given that any given GFT model defines a unique spin foam model, in its perturbative expansion. The way this link arises shows also that: 

1) the GFT prescription for completing the definition of any given spin foam model through a sum over complexes is most natural from the canonical LQG perspective, and can be motivated directly from it; 

2) to investigate aspects of the underlying canonical dynamics looking at spin foam amplitudes may not be the most convenient option, for complicated 4d models; the spin foam vertex amplitudes contains matrix elements of the full canonical projector operator, but its properties may be obfuscated in the details of the perturbative expansion and in the complexity of generic Feynman amplitudes\footnote{For the same reasons, looking at the Feynman expansion for the atomic field theory of a condensed matter system is not optimal to unravel the continuum (quantum) dynamics of a macroscopic assembly of particles}. 

In general, the established correspondence clarifies the link between LQG and spin foams and offer new field theoretic tools to study it. It shows a way (also suggested from a spin foam perspective in \cite{AlesciNouiSardelli}) to unravel the canonical dynamics underlying known spin foam, thus GFT, models, and, viceversa, to construct new spin foam/GFT models directly from any given canonical dynamics. Also, many LQG questions can now find their GFT counterpart and new field theory tools to be addressed with. In particular, the 2nd quantized GFT formalism is best suited to address issues pertaining the regime of many LQG degrees of freedom, non-perturbative (from the point of view of the spin foam expansion) and effective continuum physics \cite{GFTfluid,VincentTensor} (the recent work of \cite{GFTcondensate} is an example of progress in this direction), for the same reasons why 2nd quantization techniques are so much used in condensed matter theory.

\

The results presented here raise new questions and open new directions of investigation. Let us mention a few more of them (we have already mentioned the problem of the effective continuum geometric dynamics, the issue of constructibility and non-perturbative definition of the theory, and the possibility to construct orientation dependent spin foam dynamics via a \lq relativistic\rq ~extension of the 2nd quantized LQG formalism).

A first issue has to do with the details of the relation between the Hilbert space $\mathcal{H}$, that we have 2nd quantised and shown to give the GFT state space, and the LQG Hilbert space $\mathcal{H}_2$. In particular, it would be interesting to see if a more abstract version of the LQG Hilbert space can be defined, to bring it closer to the GFT one, and, on the other side, if any analogue of the cylindrical equivalence relations used to defined the LQG Hilbert space can be motivated and implemented in the GFT Fock space (possibly without disrupting too much the Fock structure).

One general issue is the definition and quantum geometric interpretation of the grandcanonical partition function for LQG states, allowing quantum states not solving the constraints of the canonical formulation, that we have seen corresponds to the GFT partition function. More generally still, the tentative quantum statistical setting that we have suggested for spin network states should be put on a more solid basis, with its associated thermodynamics.

This will also help, we believe, in identifying the GFT sector corresponding to the restricted sum over spin foams obtained in \cite{AlexKarim} and defining the canonical inner product of the 3d theory. This sector should have a model-independent significance and be studied also, as we anticipated, in purely field-theoretic (thus GFT) terms. 

We have seen that the natural quanta of space are open spin network vertices. We know from the canonical theory that they carry area and volume information, and we know other pre-geometric properties of them from results in quantum simplicial geometry. However, not so much has been done on the physical interpretation of open spin network links and of the non-gauge invariant degrees of freedom they carry, beside some work interpreting them in terms of matter degrees of freedom \cite{LQGmatter}. Also, at this kinematical level, what is the role of {\it colors}, such a useful addition to the GFT formalism to control the combinatorial structures it generates \cite{coloring}, from the canonical LQG perspective?

Next, the definition of the Fock space of spin networks presupposes a choice of quantum statistics. Here, we have assumed bosonic statistics, motivated as the invariance of quantum states under a discrete analogue of diffeomorphisms given by relabeling of graph vertices, thus by automorphisms of the graph itself. More work is needed to clarify the relation between continuum diffeos and discrete graph transformations \cite{LQGdiffeos} and what it implies for quantum states. In general, with the ultimate goal of a quantum geometric spin-statistics theorem for spin networks, progress should come via a careful study of symmetries to be imposed on the dynamics of spin networks, at the canonical LQG or at the GFT level. The issue is somewhat complicated by the fact that most dynamical constructions are based on the piecewise-flat category, where the status of diffeomorphism invariance is subtle. For example, the work on simplicial diffeos in GFT (and spin foams) \cite{LaurentDiffeos,GFTdiffeos} suggests a realization of diffeo invariance in terms of quantum groups, which in turn suggests a braided statistics for quantum states, in order o avoid quantum anomalies.
 
Also strictly related to the issue of symmetries, is the issue of constraints on the definition of the spin network dynamics. We have seen that in general a projector operator would have matrix elements between an arbitrary number of spin network vertices and this implies a 2nd quantized GFT dynamics  with an infinite number of interaction terms. There is nothing wrong with this, as a matter of principle, and indeed it would formally be the case for any many-body quantum theory. However, on the one hand it highlights a poor understanding on our part of what are the \lq\lq really relevant\rq\rq ~spin network interactions and the physical restrictions, e.g. coming from symmetries, that we have to impose on them; on the other hand, it is certainly impractical. This issue has been first noticed in the GFT context, the main question being what are the \lq\lq construction principles\rq\rq ~that one should follow in defining interesting dynamical models. Now we can translate it in the canonical LQG context. It has to do with the issue of symmetries, as said, and with the definition of a quantum gravity analogue of the notion of locality in standard QFT. More practically, it suggests such questions as: what is the meaning of {\it tensor invariance} introduced in the GFT context \cite{uncoloring,GFTrenorm} and related to the notion of {\it universality} of tensorial dynamics \cite{universality}? What is the quantum geometric meaning in LQG terms) of the laplacian kinetic terms introduced in the context of GFT renormalization \cite{ValentinJoseph,GFTrenorm} (and related to continuum cosmological dynamics in \cite{GFTcondensate})? Luckily, concerning these issues, the LQG/GFT correspondence also offers the tools to tackle them. In particular, the question of what are the relevant spin network interactions is not only a matter of symmetries to be imposed or of locality restrictions, but also the result of a proper analysis of the renormalization group flow of the same interactions with any intrinsic notion of scale in the theory (e.g. the modulus of the fluxes or the spins labeling quantum states). It is this renormalization flow that should tell us which of the allowed interaction terms (elementary matrix elements of the canonical projector operator) is relevant at each given scale.
 
 \
 
The same general considerations hold true for the other, many open questions this approach to quantum gravity still faces: the established correspondence between canonical LQG and GFT formalism provides a new perspective and a new set of tools to tackle them. We hope more progress will follow.

\section*{Acknowledgements}

We gratefully acknowledge the support of  the A.\ von Humboldt Stiftung through a Sofja Kovalevskaja Award. For collaboration and discussions in the early stages of this work, we are grateful to A. Youssef, E. Livine, S. Drappeau, J. Ryan and A. Di Mare. We thank M. Celoria, C. Guedes, M. Finocchiaro, A. Kegeles, T. Kittel for a critical reading of the manuscript. We are also very grateful to A. Ashtekar, S. Gielen, L. Sindoni, S. Speziale, J. Th\"urigen and, in particular, S. Carrozza, for discussions and many useful comments. Finally, the deepest gratitude is for A. Baratin, for a very careful reading of the manuscript, many detailed and important comments and crucial clarifications.

\end{document}